\documentclass[fleqn,usenatbib]{mnras}

\usepackage[T1]{fontenc}

\DeclareRobustCommand{\VAN}[3]{#2}
\let\VANthebibliography\thebibliography
\def\thebibliography{\DeclareRobustCommand{\VAN}[3]{##3}\VANthebibliography}

\usepackage{graphicx}	
\usepackage{amsmath}	
\usepackage{amssymb}	
\usepackage{multirow}
\usepackage{cuted}
\usepackage{hyperref}
\usepackage{url}
\usepackage[dvipsnames]{xcolor}
\usepackage{soul}
\usepackage{todonotes}
\usepackage{longtable}
\usepackage{pdflscape}

\usepackage[export]{adjustbox}

\newcommand{\Msun}{\mbox{$\mathrm{M}_{\sun}$}}

\newcommand{\Porb}{\mbox{$\mathrm{P}_\mathrm{orb}$}}
\newcommand {\SDSSV}{SDSS\nobreakdash-V}
\newcommand{\allsdss}{SDSS~I~to~IV}
\newcommand {\NCVs}{118 }
\newcommand {\Nperiods} {72 }
\newcommand {\Nnewperiods} {21}
\newcommand {\Nmags} {17 }
\newcommand {\Nmiscat} {eleven }
\newcommand {\Nnew} {eight }
\newcommand {\Nspec} {53 }
\newcommand {\Nprev} {110 }

\newcommand {\Nplates} {236 }
\newcommand{\mdot}{$\dot{M}$}

\newcommand\jaavso{AAVSO} 
\setuptodonotes{inline,noinlinepar,inlinewidth=2cm}
\usepackage{newtxtext,newtxmath}
\usepackage{gensymb}

\title[CVs from \SDSSV]{Cataclysmic Variables from Sloan Digital Sky Survey V -- the search for period bouncers continues}

\author[K. Inight et al.]{
K.~Inight,$^{1}$\thanks{E-mail: keith.inight@gmail.com }
Boris~T.~G\"ansicke,$^{1}$
A.~Schwope,$^{2}$
S.~F.~Anderson,$^{6}$
C.~Badenes,$^{5}$
E.~Breedt,$^{18}$
\newauthor
V.~Chandra,$^{8}$
B.~D.~R.~Davies,$^{1}$
N.~P.~Gentile~Fusillo,$^{20}$
M.~J.~Green,$^{11,1}$
J.~J.~Hermes,$^{14}$
\newauthor
I.~Achaica Huamani,$^{9}$
H.~Hwang,$^{7}$
K.~Knauff,$^{2}$
J.~Kurpas,$^{2,19}$
K.~S.~Long,$^{10}$
\newauthor
V.~Malanushenko,$^{16}$
S.~Morrison,$^{17}$
I.J.~Quiroz C.,$^{9}$
G.~N.~Aichele Ramos,$^{9}$
\newauthor
A.~Roman-Lopes,$^{3}$
M.R.~Schreiber,$^{9, 15}$
A.~Standke,$^{2,19}$
L.~St\"utz,$^{2}$ 
J.~R.~Thorstensen,$^{12}$
\newauthor
O.~Toloza,$^{9, 15}$
G.~Tovmassian,$^{4}$
N.~L.~Zakamska$^{13}$
\\
\\
$^{1}$Department of Physics, University of Warwick, Coventry, CV4 7AL, UK\\
$^{2}$Leibniz-Institut für Astrophysik Potsdam (AIP), An der Sternwarte 16, 14482 Potsdam, Germany \\
$^{3}$Departamento de Astronom\'ia, Universidad La Serena, La Serena, Chile\\
$^{4}$Universidad Nacional Aut\'onoma de M\'exico, Instituto de Astronom\'{i}a, Aptdo Postal 106, Ensenada 22860, Baja California, M\'exico\\
$^{5}$Department of Physics and Astronomy, University of Pittsburgh Allen Hall, 3941 O'Hara St, Pittsburgh PA 15260\\
$^{6}$Astronomy Department, Box 351580, University of Washington, Seattle, WA 98195, USA\\
$^{7}$Institute for Advanced Study, Princeton, 1 Einstein Drive, NJ 08540, USA\\
$^{8}$Center for Astrophysics $\mid$ Harvard \& Smithsonian, 60 Garden St, Cambridge, MA 02138. USA\\
$^{9}$Departamento de F{\'i}sica, Universidad T{\'e}cnica Federico Santa Mar{\'i}a, Avenida Espa{\~n}a 1680, Valpara{\'i}so, Chile\\
$^{10}$Space Telescope Science Institute, 3700 San Martin Drive, Baltimore, MD, 21218, USA\\
$^{11}$School of Physics and Astronomy, Tel-Aviv University, Tel-Aviv 6997801, Israel\\
$^{12}$Department of Physics and Astronomy, Dartmouth College, Hanover NH 03755, USA \\
$^{13}$Department of Physics \& Astronomy, Johns Hopkins University, Baltimore, MD 21218, USA \\
$^{14}$Department of Astronomy \& Institute for Astrophysical Research, Boston University, 725 Commonwealth Ave., Boston, MA 02215, USA\\
$^{15}$Millennium Nucleus for Planet Formation (NPF), Valpara{\'i}so, Chile\\
$^{16}$Apache Point Observatory, P.O. Box 59, Sunspot, NM 88349,USA\\
$^{17}$Department of Astronomy, University of Illinois at Urbana-Champaign, Urbana, IL 61801, USA\\
$^{18}$Institute of Astronomy, University of Cambridge, Madingley Road, Cambridge CB3 0HA, UK \\
$^{19}$Potsdam University, Institute for Physics and Astronomy, Karl-Liebknecht-Stra\ss e 24/25, 14476 Potsdam, Germany \\
$^{20}$European Southern Observatory, Karl Schwarzschild Straße 2, D-85478 Garching, Germany
}

\date{Accepted XXX. Received YYY; in original form ZZZ}

\pubyear{2020}

\begin{document}
\label{firstpage}
\pagerange{\pageref{firstpage}--\pageref{lastpage}}
\maketitle

\begin{abstract}
\SDSSV\  is carrying out a dedicated survey for white dwarfs, single and in binaries, and we report the analysis of the spectroscopy of \NCVs cataclysmic variables (CVs) and CV candidates obtained during the final plug plate observations of SDSS. We identify \Nnew new CVs, spectroscopically confirm \Nspec and refute \Nmiscat published CV candidates, and we report \Nnewperiods\ new or improved orbital periods. The orbital period distribution of the SDSS-V CVs does not clearly exhibit a period gap. In common with previous studies, the distribution shows that spectroscopically identified CVs have a larger proportion of short-period systems compared to samples identified from photometric variability. Remarkably, despite a systematic search, we find very few period bouncers. We estimate the space density of period bouncers to be $\simeq0.2\times10^{-6}\,\mathrm{pc}^{-3}$, i.e. they represent only a few per cent of the total CV population.  This suggests that during their final phase of evolution, CVs either destroy the donor, e.g. via a merger, or that they become detached and cease mass transfer.      
\end{abstract}



\section{Introduction}

Cataclysmic variables (CVs, see \citealt{2003cvs..book.....W} for a comprehensive overview) are interacting binaries in which a white dwarf is accreting from a low-mass companion. They are a key population for testing and developing our understanding of close binary evolution and interaction \citep{2023arXiv230308997B}, the shortest-period systems are among the verification sources for the space-based gravitational wave mission LISA \citep{2018MNRAS.480..302K}, and they provide ideal laboratories for the study of accretion discs \citep{2020AdSpR..66.1004H} and accretion onto magnetic stars \citep{2015SSRv..191..111F,1990SSRv...54..195C}.

However, because of their wide range of observational properties, the known population of CVs remains subject to severe selection biases \citep{2005ASPC..330....3G}. Even the census of CVs within 150\,pc is currently only $\approx77$\,per\,cent complete; the undiscovered CVs will almost entirely be systems that show little optical variability \citep{2020MNRAS.494.3799P} such as low-accretion rate WZ\,Sge systems \citep{2020PASJ...72...49T} and high-accretion rate novalike variables \citep{2022MNRAS.510.3605I}.

The Sloan Digital Sky Survey (SDSS) \citep{2000AJ....120.1579Y} has proven to be an important tool for identifying CVs spectroscopically \citep{2002AJ....123..430S,2003AJ....126.1499S,2004AJ....128.1882S,2005AJ....129.2386S,2006AJ....131..973S,2007AJ....134..185S,2009AJ....137.4011S,2011AJ....142..181S}, with the total number of CVs observed by SDSS standing at 507 \citep{2023MNRAS.524.4867I}. Follow-up studies of individual SDSS CVs led to a large number of notable individual results, including the identification of brown-dwarf donors \citep{2005ApJ...630L.173S,2006Sci...314.1578L}, the identification of low-mass transfer magnetic CVs \citep{2005ApJ...630.1037S,2007ApJ...654..521S}, the first eclipsing AM\,CVn \citep{2005AJ....130.2230A}, nuclear evolved CVs \citep{2006MNRAS.371.1435L,2014ApJ...790...28R}, pulsating white dwarfs \citep{2006MNRAS.365..969G,2007ApJ...667..433M,2014PASJ...66..113P}, and a halo CV \citep{2011MNRAS.414L..85U}, as well as the detection of spiral shocks \citep{2010ApJ...711..389A,2019MNRAS.483.1080P}. The SDSS spectroscopy led, in particular, to the discovery of a large number of CVs with no history of outbursts; analysis of these systems confirmed the long-standing prediction of a pile-up of CVs near the minimum orbital period of about 80\,min \citep{2009MNRAS.397.2170G}.

During the commissioning of SDSS \citet{2003AJ....125.2621R} developed a targeting strategy for CVs in $ugriz$ colour space, which however proved to be more effective in identifying detached white dwarf plus M-dwarf binaries. As remarkable as the impact of SDSS has been on the research of CVs, most of the SDSS CVs were observed serendipitously in a much larger pool of objects targeted for spectroscopy for a wide range of reasons, e.g. as quasar candidates or blue excess objects \citep{2002AJ....124.1810S} or because of their variability \citep{2015ApJ...806..244M}. 

SDSS has now entered its fifth phase (\SDSSV,  \citealt{2017arXiv171103234K}), which will extend multi-object spectroscopy across the entire sky by operating robotic fibre positioners on the 2.5\,m SDSS telescope at Apache Point Observatory (APO) and at the 2.5\,m Dupont telescope at Las Campanas Observatory.  \SDSSV\  differs from its predecessors in that it contains a program to  deliberately target white dwarfs and CVs. Although most of \SDSSV\  will be carried out with robotic positioners, the first eight months of \SDSSV\ observations at APO followed the approach  using fibres positioned in the focal plane with drilled plates \citep{2003AJ....125.2276B} to feed a spectrograph, as used in the earlier phases of SDSS. 

We report here on the \NCVs CVs observed in the first eight months of \SDSSV\ as part of the dedicated white dwarf binary targeting strategy.  We describe the identification of new CVs, the spectral confirmation of candidate CVs and new observations of previously known CVs. We report new orbital periods for \Nnewperiods\ of these CVs obtained by combining radial velocity data from SDSS with light curves from the Zwicky Transient Facility (ZTF, \citealt{2019PASP..131a8002B}) and other follow-up observations. The remainder of this manuscript is organised as follows.  We describe the target strategy in Section \ref{section:targeting} and our observations  in Section \ref{section:observations}.  We describe how we analyse and classify the spectra and light curves in Section \ref{section:analysis}, report our results in  Section  \ref{section:results} and discuss our findings in Section  \ref{section:discussion}.  Finally, we  summarise our results in Section \ref{section:summary}. A complete table of information about the observed systems is included in the Appendix.

\section{Target selection}\label{section:targeting}
Whereas the previous generations of the SDSS surveys were prolific in identifying new CVs, the vast majority of these systems were observed serendipitously, many of them because they overlap in colour space with quasars \citep[see for example][]{2002AJ....123.2945R,2009MNRAS.397.2170G}. \allsdss preceded \textit{Gaia} \citep{2018A&A...616A...1G}, therefore parallaxes, and hence absolute magnitudes were not available, and the only dedicated target selection for CVs relied on rather crude colour-colour cuts, which were in fact more successful in identifying detached white dwarf plus M-dwarf binaries \citep{2003AJ....125.2621R}. The CVs within \allsdss were identified by visual inspection of the SDSS spectroscopy, sometimes aided by various techniques filtering the entire SDSS data set into shortlists of candidate objects to be inspected \citep[e.g.][]{2002AJ....123..430S,2003AJ....126.1499S,2004AJ....128.1882S,2005AJ....129.2386S,2006AJ....131..973S,2014MNRAS.439.2848C,2015MNRAS.448.2260G,2019MNRAS.486.2169K,2021MNRAS.507.4646K,2023MNRAS.524.4867I}. 

\SDSSV\ provided the first opportunity to define dedicated target selections aimed at white dwarfs and CVs that made use of the \textit{Gaia} astrometry. These target selections are defined using ``cartons'', most of which use a set of algorithmic rules, filtering large catalogues of objects, to produce the target lists for \SDSSV\  (see Table\,2 in \citealt{2023arXiv230107688A} and the description of the \href{https://www.sdss.org/dr18/mwm/programs/cb/}{compact binary} and \href{https://www.sdss.org/dr18/mwm/programs/wd/}{white dwarf cartons}). We defined eight cartons using a range of criteria to target white dwarfs, both single and in binaries: \texttt{mwm\_wd\_core} is based on the \textit{Gaia} white dwarf candidate catalogue of \citet{2019MNRAS.482.4570G}, whereas \texttt{mwm\_cb\_gaiagalex, mwm\_cb\_uvex1, mwm\_cb\_uvex2, mwm\_cb\_uvex3, mwm\_cb\_uvex4 and mwm\_cb\_uvex5} leverage the fact that both single white dwarfs and those in binaries (detached and CVs) have an ultraviolet excess. These ultraviolet-excess cartons are based on cross-matching \textit{Gaia} with ultraviolet surveys including \textit{GALEX} \citep{2007ApJS..173..682M}, the ultraviolet  observations obtained by the Optical Monitor of \textit{XMM-Newton} \citep{2001A&A...365L..36M,2012MNRAS.426..903P} and the UVOT instrument onboard \textit{SWIFT}  \citep{2004ApJ...611.1005G,2005SSRv..120...95R}. These ultraviolet-excess cartons contain a substantial proportion of, but not all, the CVs in the observable \SDSSV\ footprint. Finally, we defined a carton, \texttt{mwm\_cb\_cvcandidates} that contains a collection of published  CVs and CV candidates (potential CVs that have typically been identified by an outburst and need confirmation with a spectrum). Full details of the target selection rules will be provided in a forthcoming publication. These eight cartons were then assigned priorities for the \SDSSV\  observations that were above most of the other stellar and extragalactic cartons, ensuring a high completeness of the spectroscopic follow-up. One important fact to bear in mind in the analysis of the \SDSSV\  results is that a given target can be selected by multiple cartons.

When referring to individual systems, we abbreviate their SDSS designations to four digits in each of RA and Dec, e.g. SDSS\,J062429.71+002105.8 is referred to as J0624+0021. The full designations are listed in Table\,\ref{tab:cv_master_table}.

\section{Observations}\label{section:observations}
\begin{figure} 
\includegraphics[width=\columnwidth]{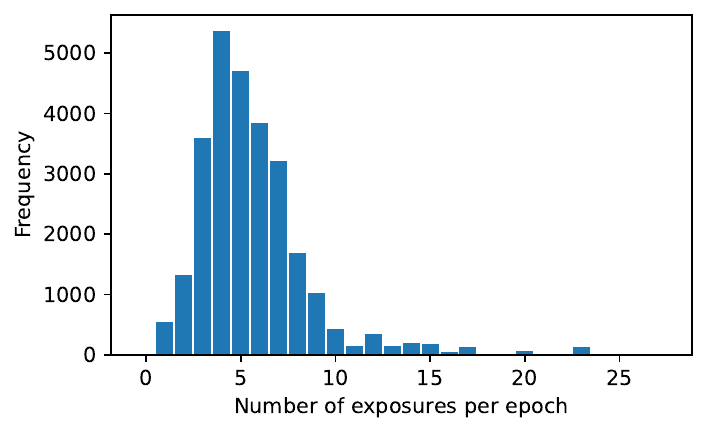}
\caption{\label{fig:Histofnumexp} Distribution of the number of 900\,s exposures used to form each of the 27191 co-added spectra. The exposures were often contiguous but always obtained within a 2--3 day epoch. Where an object was observed in more than one epoch a separate co-added spectrum was produced for each epoch}
\end{figure}

\begin{table}
\caption{\label{tab:SDSSV_observations}  
Target cartons observed by \SDSSV\  and used for the analysis in this paper showing the completeness. The numbers observed per carton refer to the \Nplates plates observed as part of \SDSSV\  in which these cartons were implemented in the observing strategy. Many targets are present in multiple cartons.}
\begin{tabular}{lrrrr}
\hline
Carton & Targets & Number & per\,cent & CVs\\
 &  & observed & complete & found \\\hline
mwm\_cb\_gaiagalex & 7773 & 4050 & 52 & 26 \\
mwm\_cb\_uvex1 & 3281 & 2488 & 76 & 27 \\
mwm\_cb\_uvex2 & 8040 & 5802 & 72 & 44 \\
mwm\_cb\_uvex3 & 29 & 24 & 83 & 0 \\
mwm\_cb\_uvex4 & 179 & 144 & 80 & 2 \\
mwm\_cb\_uvex5 & 282 & 27 & 10 & 0 \\
mwm\_wd & 6229 & 5781 & 93 & 18 \\
mwm\_cb\_cvcandidates & 141 & 124 & 88 & 101 \\ \hline
 
\end{tabular}
\end{table}

\subsection{SDSS Spectroscopy} \label{sec:SDSS_spectroscopy}
The SDSS BOSS spectrograph \citep{2013AJ....146...32S,2013AJ....145...10D}  covers the range  $3600-10\,400$\,\AA. The data analysed here was processed with v6\_1\_0  of the \SDSSV\  pipeline \citep{2023arXiv230107688A}, which performed sky subtraction together with flux and wavelength calibration of each exposure. Based upon our analysis the \SDSSV\  spectroscopy can be used to measure radial velocities from narrow spectral lines with an accuracy of up 20\,km\,s$^{-1}$. 
\SDSSV\  differs from the preceding surveys in that it expanded the coverage to include the Galactic disc, which introduces new challenges; in particular the reddening effect of interstellar dust has impacted the flux calibration of a number of spectra.

\SDSSV\  spectroscopy of the cartons described in Sect.\,\ref{section:targeting} was carried out on \Nplates individual plates, where each plate covers $\simeq7\,\mathrm{deg^{2}}$ of the sky. 
Sixteen pairs of plates had identical plate centre coordinates. The number of targets per carton, observed targets per carton, spectroscopic completeness, and the number of CVs identified per carton are reported in Table\,\ref{tab:SDSSV_observations}. The exposure time for each plate observation was 900\,s. A number of exposures were taken which were often, but not always, contiguous. The pipeline collects all the exposures of a given plate within $2-3$\,d (a ``plate-epoch'' hereafter referred to as an epoch) and co-adds the individual spectra for each target (see Fig.\,\ref{fig:Histofnumexp}). We use the co-added spectrum for our spectral classification identification as it will have a higher signal-to-noise ratio (SNR) than the individual 900\,s spectra. We use the individual spectra to probe for variations in radial velocity. The plate observations of \SDSSV\  collected 27\,191 co-added spectra of 11\,384 individual targets that fall within our cartons (many plates were observed at different epochs, separated by weeks to months, resulting in multiple co-added spectra of a substantial number of targets). 

The CV sample presented in this paper is therefore not complete within the \SDSSV\ plate program and is limited to the systems targeted in the cartons described above~--~additional CVs may emerge from the extragalactic programs, although these are most likely to be too faint to have a \textit{Gaia} counterpart.

\subsection{Photometric survey data}
We retrieved archival light curves from the Catalina Real-Time Transient Survey (CRTS,  \citealt{2011arXiv1102.5004D}), Zwicky Transient Facility (ZTF, \citealt{2019PASP..131a8002B})  and the Transiting Exoplanet Survey Satellite (\textit{TESS}, \citealt{2015JATIS...1a4003R}). We reviewed these primarily for historical outbursts. The ZTF and \textit{TESS} light curves are shown alongside the spectra in  Figs.\,1 to 13 in the supplementary material.

\subsection{Follow-up observations}
\subsubsection{Photometry}
We obtained follow-up photometry using the 2\,m Liverpool Telescope (LT, \citealt{2004SPIE.5489..679S}, see Table\,\ref{tab:LTl_observations} for details)  of two particularly interesting CVs: J0624+0021, which is located in the period gap (Sect.\,\ref{sec:j0624}) and J1740+0258 which exhibits unusual state changes (Sect.\,\ref{sec:j1740}). Both systems were identified as potentially eclipsing systems from their SDSS spectra where the higher Balmer lines exhibit deep central absorption dips that go down to, or even below the continuum.  Both targets were observed twice. The length of the initial observation was chosen to exceed the likely orbital period so that at least one eclipse would be observed. In both cases, the first observation covered two eclipses, from which we estimated the orbital period, and then scheduled a second observation a few days later to provide a more accurate period measurement. Each LT observation consisted of a sequence of 90\,s exposures with the IO:O imager using a Bessel-$V$ filter. The standard LT pipeline was then used to provide bias subtraction and flat fielding. Differential photometry was obtained with the \textsc{SExtractor} package \citep{1996A&AS..117..393B}, using \textit{Gaia} EDR3 3120280993084993792 as comparison star for  J0624+0021 and \textit{Gaia} EDR3 4376322089783941120 for J1740+0258.

\begin{table}
\caption [.] {\label{tab:LTl_observations}  
The journal of the Liverpool Telescope observations.}
\begin{tabular}{|l|l|l|l|}
\hline
Date & UTC&$N$ (obs)&Duration\\
& (start)  & &(minutes)\\ \hline
J0624+0021& & & \\
2021 Dec 29 & 22:27:40 & 120 & 218 \\
2022 Jan 05 & 22:40:53 & 90 & 162 \\ \hline
J1740+0258 & & & \\
2022 Jun 03 & 23:09:51 & 180 & 325 \\
2022 Jun 20  & 22:41:14 & 200 & 262 \\ \hline
\end{tabular}
\end{table}

\subsubsection{Spectroscopy}
Prior to \SDSSV, one of the authors (JRT) obtained time-resolved spectroscopy using the 2.4\,m Hiltner telescope of the MDM observatory at Kitt Peak (see \citealt{2020AJ....160....6T} for more details on the instrument and data reduction techniques) of the CV candidate J0418+5107 (NS\,Per) because it appeared to be suitable for a radial-velocity based period determination. We present here the so far unpublished results of these observations (Sect.\,\ref{sec:NSPER}).

\section{Analysis}\label{section:analysis}

\subsection{Classification}
Classification consists of firstly identifying the CVs among the 27\,191 \SDSSV\  co-added spectra and then determining the sub-type of each CV.

\subsubsection{CV identification}\label{section:identification}
We initially scanned each spectrum by eye for indications of a CV nature~--~typically the presence of emission lines but see \citet{2023MNRAS.524.4867I} for a full description of the spectral characteristics of CVs. Although in theory this process could be automated there was a risk that some important exotic object could be missed (such as novalike variables and dwarf novae observed during outburst, which have disc-dominated spectra with, in some cases, very weak absorption lines). In addition some plates near the Galactic plane that were subject to substantial amounts of reddening had very poor flux calibrations, and CV spectra obtained on those plates would very likely have been missed in any automated search. 

The list of CVs obtained from this visual inspection was then compared with CV catalogues \citep{2003A&A...404..301R,2017yCat....102027W} to identify previously known CVs and then subjected to more detailed scrutiny using the available archival information (light curves, astrometric and photometric data). New CVs identified in \SDSSV\  and systems where the \SDSSV\  spectra refute previously published CV classifications are discussed in Sect.\,\ref{section:results}, detailed notes on previously known CVs where the \SDSSV\  spectra confirms their classification are given in Appendix\,\ref{section:prevknown} and Appendix\,\ref{section:wellknown} includes notes on a selection of well known CVs where \SDSSV\  has revealed new information. 

Several CVs were observed at different  epochs by \SDSSV\  and the resulting co-added spectra were compared and merged where appropriate. As part of this process we found four CVs (J0528$-$0333, J0808+3550, J0926+0345, J1830+2655) that were observed in both outburst and quiescence and these are shown separately in Fig.\,13 in the supplementary material. 

The final list of CVs (Table\,\ref{tab:cv_master_table}) was then subjected to a rigorous search of the literature to find references both for the initial claim of a CV nature, the first published spectrum, and the most accurate measurement of the orbital period. 

\subsubsection {CV sub-types}
CVs are categorised into a number of sub-types based on their observed characteristics, with the sub-types often being named after a prototype CV that exemplifies particular characteristics. In this paper we have used the classifications from the literature where available, making additions and corrections where appropriate. 

In the companion paper on CVs observed by in \allsdss\ \citep{2023MNRAS.524.4867I} we describe the taxonomy of CVs and summarise the salient stages of CV evolution. The following is a highly simplified overview:

Non-magnetic CVs ($B\lesssim1$\,MG) that exhibit disc outbursts are dwarf novae, those with steady hot discs are novalike variables. Dwarf novae are further sub-classified into SU\,UMa CVs (typically short-period, $\Porb\lesssim3$\,h, that have relatively frequent outbursts, interspersed by longer and brighter superoutbursts), ER\,UMa CVs (systems with very short superoutburst recurrence times), WZ\,Sge CVs (which only have rare superoutbursts), U\,Gem CVs (typically long-period, $\Porb\gtrsim3$\,h, dwarf novae that do not show superoutbursts), and Z\,Cam CVs (that switch between outburst states and ``stand-still'' periods of constant brightness). CVs with highly magnetic white dwarfs (MCVs) consist of polars (which typically have $B\gtrsim10$\,MG and the white dwarf spin period synchronised with the orbital period, $P_\mathrm{spin}=\Porb$) and intermediate polars (IPs, with $10\gtrsim B\gtrsim1$\,MG, and $P_\mathrm{spin}<\Porb$). Finally, AM\,CVn CVs are ultra-short period ($\Porb\lesssim60$\,min) hydrogen-deficient CVs. 

We classified all \SDSSV\  CVs by considering all available data. This includes not only the spectrum but also CRTS, ZTF and \textit{TESS} light curves, spectral energy distributions (SED) based on the available broad-band photometry, Hertzsprung-Russell diagrams based on the \textit{Gaia} astrometry and photometry, and Pan-STARRS images.  CVs for which we were unable to determine a sub-type are simply classified ``CV''.

\subsection{Orbital Periods}
We use a combination of spectroscopically derived radial velocities (SDSS and the MDM Observatory) and photometric light curves (ZTF) to estimate orbital periods.   

\subsubsection {Radial velocities from \SDSSV}
The measurement of radial velocities from the Doppler shifts of emission lines, and their variation over an orbital period, has a long history of determining the period of a CV \citep{1923ApJ....58..215M}. The emission lines are assumed to arise from the accretion disc around the white dwarf and to track the orbit of the white dwarf around the centre of gravity of the CV. The observed velocity is smaller than the Keplerian velocity of the white dwarf, $v_\mathrm{obs}=\sin i \times v_\mathrm{wd}$ with $i$ the inclination of the binary, which implies that orbital periods for low inclination systems may be difficult to impossible to measure. 

\begin{figure*} 
\includegraphics[width=\textwidth]{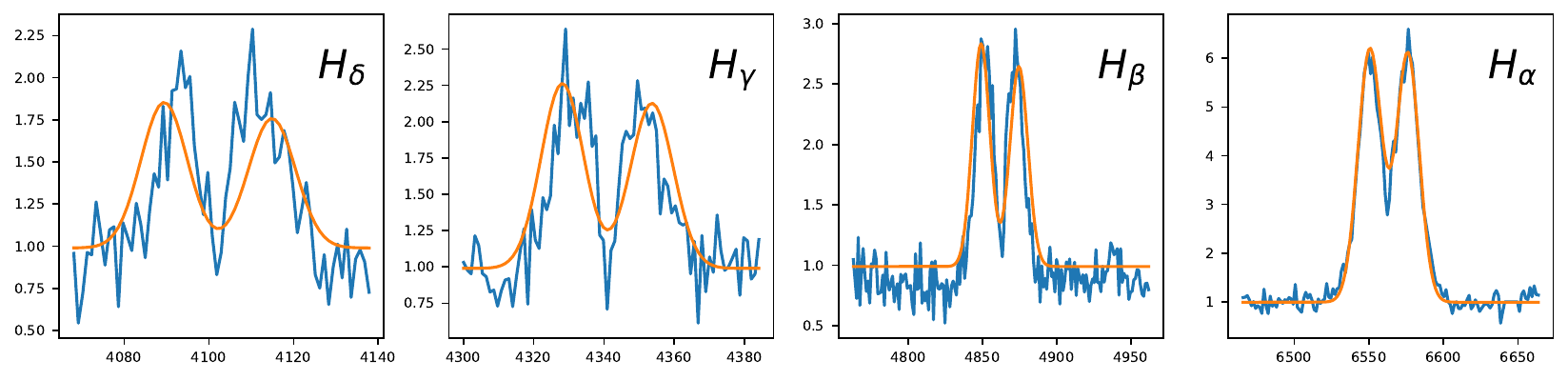}
\caption{\label{fig:RVtemplate} Example of extracting radial velocities by fitting a template of four double-Gaussian profiles (orange) to an individual 900\,s spectrum of J0633+0303 (blue). The observed spectrum and the template are both normalised to unity individually for each Balmer line. The radial velocity of the CV is measured by fitting all four Balmer lines simultaneously with a single free parameter for the velocity.}
\end{figure*}

The spectroscopy of the CVs in \SDSSV\  consists of one or more epochs in each of which multiple 900\,s exposures were obtained (see Fig.\,\ref{fig:Histofnumexp}). In contrast to the co-added identification spectra shown in Sect.\,\ref{section:results} and the supplementary material, many of the individual 900\,s exposures have low SNR, and some of them are affected with cosmic ray artefacts that had to be manually identified and removed. In order to measure radial velocities from these individual spectra, we developed a two-step procedure simultaneously fitting two Gaussians (to model double-peaked emission lines) to each of the first four Balmer lines, H$\alpha$ to H$\delta$,  using the same radial velocity (Fig.\,\ref{fig:RVtemplate}). The \textsc{lmfit} package provides a straightforward way to achieve this in \textsc{python}. Prior to fitting we normalised the continuum around each of the four Balmer lines in each exposure to unity by means of a first-order polynomial fit. 

In the first step we notice that a Gaussian has three free parameters~--~central wavelength ($\mu$), width ($\sigma$) and amplitude ($A$):
\begin{equation}
f(\lambda;A,\mu,\sigma) = \frac{A}{\sigma \sqrt{2\pi}} \exp\left(-\frac{1}{2}\left(\frac{\lambda-\mu}{\sigma}\right)^{2}\right) 
\end{equation}
For this step we created an \textsc{lmfit} \citep{2016ascl.soft06014N} model of the form:
\begin{multline} \label{eq:lmfit}
C+\sum_{i \in \{\ \alpha, \beta, \gamma, \delta \}}^{} \Bigl(
f\left(\lambda;A_{i,1},\lambda _{i} -s/2,\sigma_{i,1}\right) \Bigr.\\ + \Bigl. f\left(\lambda;A_{i,2},\lambda_{i} +s/2,\sigma_{i,2}\right) \Bigr) 
\end{multline}
where
\begin{equation}
\lambda_i= \mu_i+\sqrt{\frac{1-v/c}{1+v/c}}
\end{equation}
Here $\mu_i$ is the central wavelength of the i(th) Balmer line, $s$ is the difference  in wavelength between the centres of the two Gaussians and $v$ is the radial velocity causing the Doppler shift; $C$ is a constant. This model is fitted to each exposure in turn allowing all the free parameters to vary. In a few cases where the spectrum shows a steep Balmer decrement and the higher order lines are not visible we limit the process to H$\alpha$ and H$\beta$. 

For the second step a template model of the form (\ref{eq:lmfit}) is formed by averaging  the values of each of the free parameters derived in the first step over the exposures. The template model is then fitted to each exposure with only $v$ being allowed to vary.

With a set of measured radial velocities in hand, the next step is to estimate the period. This is inherently difficult because the dataset is sparse and so we use a two (three) pronged approach, based on sine fits to the radial velocities, the use of the   \textsc{Python} package called \textsc{The Joker} \citep{2017ApJ...837...20P}  and ZTF periodograms where available. We consider a period trustworthy where the approaches give consistent results. To fit a sine wave we use \textsc{lmfit} to fit an equation of the form:
\begin{equation}
g(t;K,P,\phi,v_\mathrm{sys}) = K \sin\left(\frac{2 \pi t}{P} +\phi\right)+v_\mathrm{sys}
\end{equation}
where $K$, $P$ and $\phi$ are the amplitude, period, and phase of the radial velocity variation, and $v_\mathrm{sys}$ is the systemic velocity. Unfortunately when seeking the optimal values of the four free variables  \textsc{lmfit} is very sensitive to the initial values with which it is seeded and will select a solution giving a local minimum for $\chi^2$ which will not always be the correct one. To address this problem, we used an existing published value or a photometric estimate (e.g. from ZTF) where available to seed the process. 

We also used  \textsc{The Joker}  to help identify the correct alias. \textsc{The Joker} makes prior assumptions about the distribution of six parameters~--~namely orbital period, eccentricity\footnote{For the purpose of our analysis, we assume that the orbit is circular and that any small ellipticity can be ignored.}, pericenter phase and argument, velocity semi-amplitude and the barycenter velocity. These distributions are used  to create a large set of samples which are then compared with the observed data to obtain posterior distributions and hence the most likely alias.   

The accuracy of a  period derived from radial velocities depends heavily upon the SNR of the data, the number of exposures and the number of epochs.

\subsubsection{Light curves from ZTF}
\begin{figure} 
\includegraphics[width=\columnwidth]{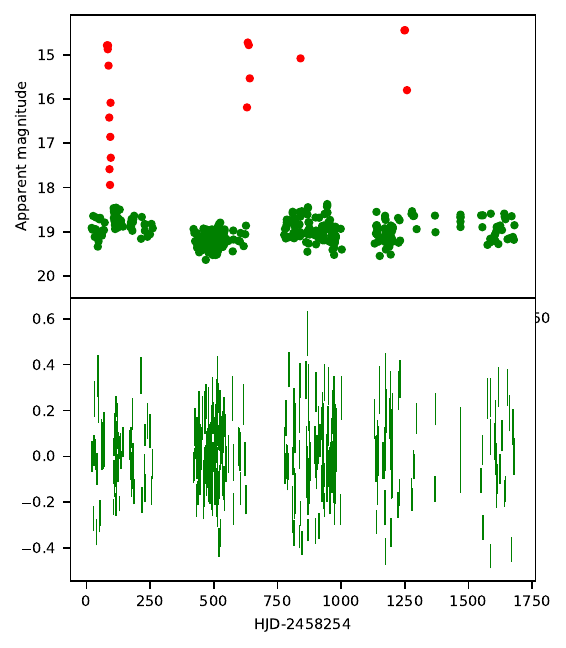}
\caption{\label{fig:Fourieranalysis1} Top panel:  ZTF light curve of J0038+2509 showing the outbursts in red. Bottom panel: The same light curve, clipped to remove the outbursts and with long-term trends removed.}
\end{figure}

\begin{figure} 
\includegraphics[width=\columnwidth]{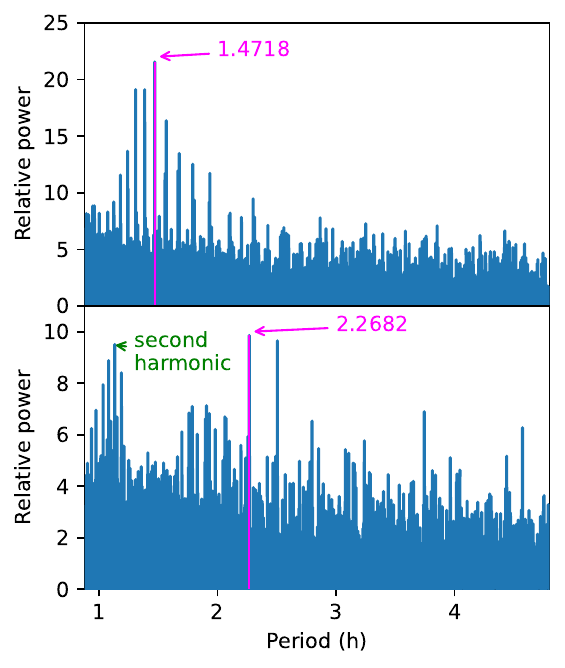}
\caption{\label{fig:Fourieranalysis2} Top panel: The ZTF power spectrum of J0635+0303. The strongest signal is likely to be the orbital period and was used to seed the analysis of the radial velocity variations. Bottom panel: The ZTF power spectrum of J0038+2509, showing the strongest signal at 2.26\,h, and a slightly weaker signal at the second harmonic. The strongest signal is consistent with the superhump period of 2.33\,h found by \citet{2012PASJ...64...21K} and is therefore the orbital period.}
\end{figure}

The plate survey of \SDSSV\   and ZTF have similar coverage in the northern sky, so most of our CVs and CV candidates have ZTF data. ZTF saw first light in November 2017 and is designed to survey the sky every $\simeq3$\,d, although some areas are observed more frequently in the search for extragalactic transients \citep{2019PASP..131a8002B}. The number of ZTF photometric data points varies with the location on the sky with median values of 260 and 482 for the $g$ and $r$-band, respectively. Some objects additionally have a few nights of high cadence (with images obtained typically every few minutes for several hours) observations, which are particularly useful for measuring orbital periods. Before searching for periodic behaviour it is necessary to remove outbursts and also remove any long term trends (see Fig.\,\ref{fig:Fourieranalysis1}). We defined a threshold for each object to clip outbursts and used a polynomial filter to remove trends. The filter spans a longer time than any potential orbital period. We then computed a periodogram (see Fig.\,\ref{fig:Fourieranalysis2}) using two techniques:  a Fourier analysis \citep{1975Ap&SS..36..137D} and the ``Multi-harmonic Analysis of Variance'' (AOV) algorithm \citep{1989MNRAS.241..153S}. The Fourier analysis works well for sinusoidal light curves whilst the AOV is superior for light curves with more complex structures, e.g. cyclotron beaming or eclipses.  For the 66 systems studied here that had no period measurement,  ZTF periodograms provide a unique reliable period in only eight cases (typically where there is high cadence data) but they also  help to identify the correct alias from periods derived from radial velocities. Depending on the physical origin of the photometric modulation, the strongest signal in the ZTF periodograms could be equal to the orbital period (e.g. for eclipsing CVs) or half the orbital period (e.g. for systems where the donor contributes significant amounts of light, resulting in ellipsoidal modulation, or cyclotron beaming in a polar), so care has to be taken in the interpretation of photometric periods.   

\subsubsection{Eclipses from the Liverpool Telescope}
We measured the eclipse timings of J0624+0021 and J1740+0258 from the LT light curves (Fig.\,\ref{fig:LT_telescope} and Fig.\,\ref{fig:LT_telescope_2}) by fitting a Gaussian to the eclipse profiles. In each case, two eclipses were covered by the initial LT observation, and accurate periods were then determined from this estimate combined with the second sets of LT data, and the periodogram calculated from the ZTF data. 

\section{Results}\label{section:results}
Details of all \NCVs CVs observed in \SDSSV\  are listed in Table \ref{tab:cv_master_table}. We identified \Nnew new CVs, obtained the first spectrum for 45 previously known CVs and CV candidates, and disproved the published CV classification of \Nmiscat systems. The co-added spectra and light curves for the \Nnew new CVs are shown in Fig.\,\ref{fig:NewCVSpecLC01} and the properties of these systems are discussed in Sect.\,\ref{section:newCVs}. The spectra and light curves of \Nprev previously known CVs and candidate CVs are shown in Figs.\,1 to 13 in the supplementary material and any new information on these systems is discussed in Appendices\,\ref{section:prevknown} and \ref{section:wellknown}. The spectra and light curves of the \Nmiscat non-CVs are shown in Figs.\,\ref{fig:NotsCVSpecLC01} and  \ref{fig:NotsCVSpecLC02}, and we discuss the likely nature of these systems in Sect.\,\ref{sec:misclassified}. Lastly, an additional 13 published CV candidates were observed by \SDSSV, but their spectra were unusable (Table\,\ref{tab:unusable_observations}).

\subsection{New CVs}\label{section:newCVs}
\begin{figure*} 
\includegraphics[width=0.99\textwidth]{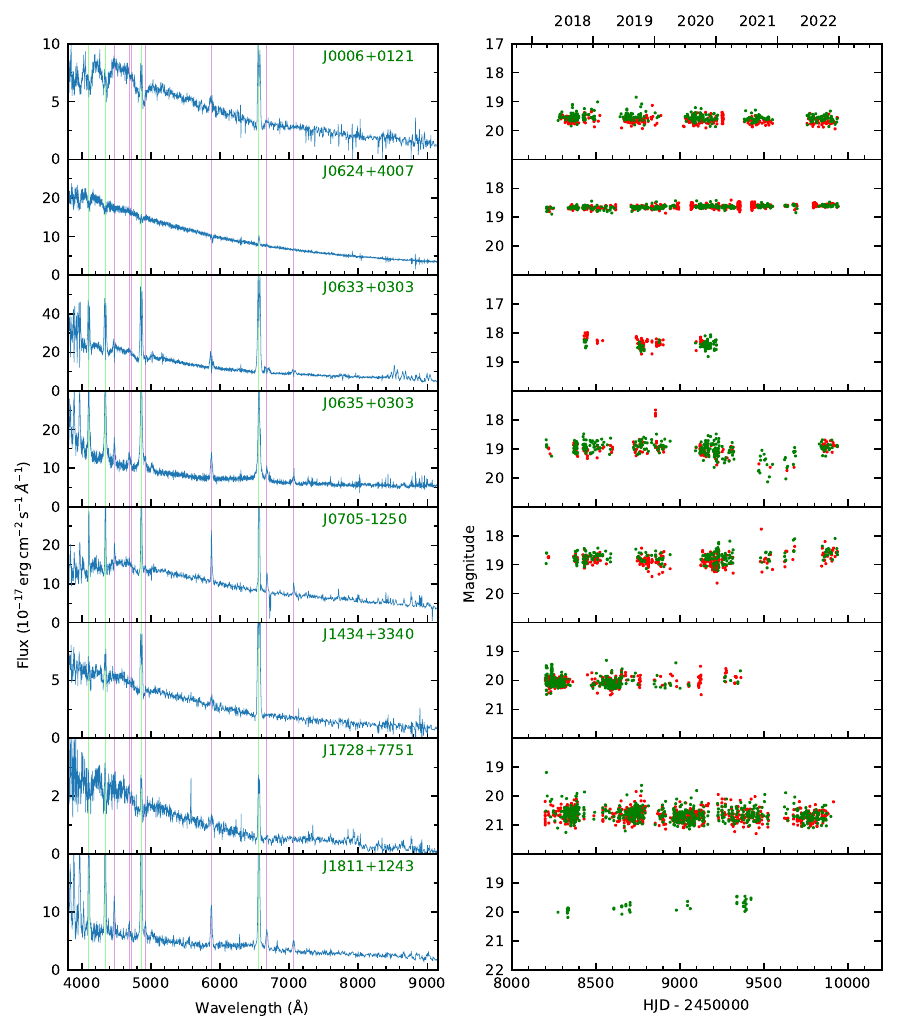}
\caption{\label{fig:NewCVSpecLC01}
Left panels: Spectra of the eight new CVs discovered by SDSS-V.
Right panels: ZTF light curves  ($r$- and $g$-band) of the eight new CVs.}   
\end{figure*}

Inspection of Fig.\,\ref{fig:NewCVSpecLC01} shows that amongst the ZTF light curves of the eight new CVs there is only one outburst (J0635+0303) which is very likely a reason why these systems have so far escaped attention.

\subsubsection{J0006+0121} 
The spectrum shows strong H$\alpha$ emission, and a steep decrement in the higher lines of the Balmer series. The double-peaked morphology of H$\alpha$ is evidence of an accretion disc seen at moderately high orbital inclination. The continuum rises towards the blue which, together with the broad H$\gamma$ and H$\delta$ absorption lines, is interpreted as being dominated by the emission of the white dwarf. This assumption is consistent with the location of J0006+0121 within the white dwarf cooling sequence. The fact that the white dwarf is visible, combined with the absence of outbursts in the CRTS and ZTF light curves, indicates a low mass transfer rate.  No spectroscopic features of the donor star are detected in the red part of the spectrum and the weak infrared peak in the SED plot is suggestive of a brown dwarf companion. We classify this as a WZ\,Sge. 

A periodogram computed from the radial velocities measured from the eight individual SDSS spectra displays a number of possible periods separated by one-day aliases. Whereas the SDSS spectroscopy is not sufficient to unambiguously identify the orbital period, the most likely values from  \textsc{The Joker} periodogram are 1.434(3)\,h, 1.525(4)\,h, and 1.628(4)\,h, where the uncertainties have been determined from sine fits to the data. We were unable to obtain a period from either the CRTS or ZTF data.  If either of the two longer periods represent the true orbital period, this system would likely be a  ``period bouncer''.

\subsubsection{J0624+4007}
The spectrum exhibits a steep blue continuum and weak, narrow H$\alpha$ emission line. The higher Balmer lines display narrow absorption profiles (with a hint of an emission core in H$\beta$). The ZTF light curve shows the system at constant brightness. With $G_\mathrm{abs}=7.35$ and $G_\mathrm{BP}-G_\mathrm{RP}=0.39$ this CV is located near the bottom of the area of the HR diagram populated by novalike variables (see Fig.\,6 in \citealt{2022MNRAS.510.3605I}). Combining all observational evidence suggests that J0624+4007 is a novalike. The relatively faint absolute magnitude suggests that it is either a high-inclination system (i.e. nearly edge-on) or, taking into account the absence of eclipses, one with a relatively low mass transfer rate, or both. We were unable to obtain an orbital period from either the radial velocities or the ZTF photometry. 

\subsubsection{J0633+0303}
The spectrum shows strong double-peaked Balmer and \ion{He}{i}  emission lines on a red continuum. However, the flux calibration of this spectrum is problematic (see Sect.\,\ref{sec:SDSS_spectroscopy}), as the system is located in the Galactic plane ($b=-2.7\degree$) with a maximum reddening of $E(B-V)\simeq1.1$ along this line of sight \citep{2011ApJ...737..103S}. However, with a distance of $\simeq242$\,pc,

the system is in front of most of the dust \citep{2019A&A...625A.135L}, and inspecting the SED plot of J0633+0303 reveals that it is a blue object. With that knowledge in mind, the spectrum reveals broad depressions suggesting that the white dwarf dominates the emission, which is consistent with the position near the white dwarf cooling sequence in the \textit{Gaia} HR diagram, and is indicative of a low level of accretion. Analysis of the SDSS-V radial velocities indicates a period of approximately 1.4\,h, however the sampling  of the available spectroscopy results in several possible aliases near 1.4\,h so that we are unable to  unambiguously identify the correct period. The ZTF light curve is relatively sparse, and contains no outbursts. The very deep central absorption in the Balmer lines suggests a high inclination, and possibly eclipsing nature of J0633+0303. Based on the spectroscopic appearance, period and absence of outbursts, it is most likely a WZ\,Sge dwarf nova.

\subsubsection{J0635+0303}
The spectrum shows strong Balmer and \ion{He}{i} emission lines. The  \ion{He}{ii} line is very weak. The asymmetric morphology of the emission lines, and their very broad wings suggest that J0635+0303 is a magnetic CV, although the spectrum does not reveal any cyclotron humps. J0635+0303 is located in the Galactic plane ($b=-2.2\degree$)  with a maximum reddening of $E(B-V)\simeq1.5$  resulting in bad flux calibration of the SDSS spectrum. However, with a distance of $\simeq643$\,pc, the CV will only be reddened by $E(B-V)\simeq0.1$ and, as is evident from the broad-band photometry, it is intrinsically blue. J0635+0303 has been detected as an X-ray source in a 1.4\,ksec exposure \citep{2020ApJS..247...54E}. The ZTF light curve contains a ${\Delta m\simeq1.3}$ outburst  at $\mathrm{MJD}=58\,854$, and a dip in its brightness between $\mathrm{MJD}\simeq59\,200$ and 59\,800. The ZTF power analysis shows a strong peak at 1.47\,h. There are three epochs of good spectroscopic data and using the photometric period as a seed resulted in a period of 1.4717\,h. 
Visual inspection confirms that $\Porb=1.4717(2)$\,h is the most likely orbital period. We conclude that J0635+0303 is probably a polar.


\subsubsection{J0705--1250}
The spectrum shows strong  Balmer and \ion{He}{i} emission lines and broad Balmer absorption lines from the white dwarf, but no sign of the donor. The emission lines are narrow indicating a low inclination. The ZTF light curve does not contain any outbursts. Although there are \textit{Swift} X-ray detections this system is close (246\,pc) and so the detection of X-rays is not necessarily suggestive of it being a magnetic CV. J0705--1250 is located on the white dwarf cooling sequence in the HR diagram.   Analysis of the radial velocities measured from five epochs of reasonably good radial velocity data reveals a best-fit period of 1.4986(1)\,h. This value is, however, not consistent with ZTF periodograms which reveal potential periods of 2.34\,h and 2.13\,h. Given the somewhat mixed set of characteristics, we classify J0705--1250 as a generic CV, and we encourage follow-up studies of this system to accurately determine the orbital period, and stringently rule out a magnetic nature of the white dwarf. 

\subsubsection{J1434+3340}
This object was identified as an optical counterpart to \textit{Chandra} CXOXB\,J143435.3+334048 by \citet{2006ApJ...641..140B} who were searching for AGNs. It has double peaked Balmer and \ion{He}{i}  lines but no evidence of \ion{He}{ii} at $4686$\,\AA. The white dwarf absorption lines are visible, but there is no signature of the donor in either the spectrum or the SED. J1434+3340 sits on the white dwarf cooling sequence in the HR diagram.  We could not recover any radial velocity variation. There are no outbursts in the CRTS and ZTF light curves, and the photometric data does not reveal any periodicity. We conclude that this is a WZ\,Sge  and would expect it to have a period close to the period minimum.

\subsubsection{J1728+7751}
The spectrum shows double-peaked Balmer and \ion{He}{i} emission lines on top of a blue continuum. The white dwarf absorption lines are visible but there is no spectroscopic detection of the donor. The ZTF light curve shows a single bright outlier which seems insufficient to qualify as an outburst detection.  We could neither recover any radial velocity variations from the \SDSSV\  spectra,  nor could we derive a period from the ZTF data. J1728+7751 is located close to the white dwarf cooling sequence in the HR diagram, and it is probably a WZ\,Sge.

\subsubsection{J1811+1243}
The spectrum shows strong Balmer, \ion{He}{ii} and \ion{He}{i} emission lines, and there is no sign of either the white dwarf or the donor. The absence of a spectroscopic signature of the donor suggests that the system is below the period gap.  J1811+1243 is closer to the white dwarf cooling sequence than the main sequence in the HR diagram, indicating a relatively low accretion rate. The ZTF light curve contains  only 32 observations; there are no CRTS observations. We have seven radial velocity measurements over three epochs of varying SNR but cannot derive a reliable period. We conclude that this is probably a SU\,UMa.

\begin{table*}
\caption{Previously known CVs and CV candidates which have been misclassified in the literature.} \label{tab:notacv}

\begin{tabular}{lll}
\hline
SDSS name & Alternative name & New category \\ \hline
SDSS\,J012156.25+143737.2   & MLS160708:012156+143737 & Extra-galactic \\
SDSS\,J043017.85+360326.6   & MASTER OT J043017.84+360326.9 & YSO\\
SDSS\,J061043.98+251031.0   & NSV 2853 & Star \\
SDSS\,J064939.16$-$060005.0   & BEST-II lra2b\_01098 & Detached binary\\
SDSS\,J071844.74$-$242546.3   & Gaia18ajg & YSO: \\
SDSS\,J080826.00+314125.3   & MASTER OT J080826.00+314125.6 & Extra-galactic\\
SDSS\,J121929.32+471522.9   & SDSS\,J121929.46+471522.8 & White dwarf \\
SDSS\,J132444.33$-$142335.7   & Gaia17aoi & White dwarf\\
SDSS\,J162450.04+654101.1   & Gaia16aat & Quasar\\
SDSS\,J165449.45+350803.7   & MLS160613:165449+350804 & Extra-galactic\\
SDSS\,J195919.93+162440.3   & Gaia17bqf & Star \\\hline
\end{tabular}
\end{table*}

\subsection{Misclassified systems}\label{sec:misclassified}
\begin{figure*} 
\includegraphics[width=0.99\textwidth]{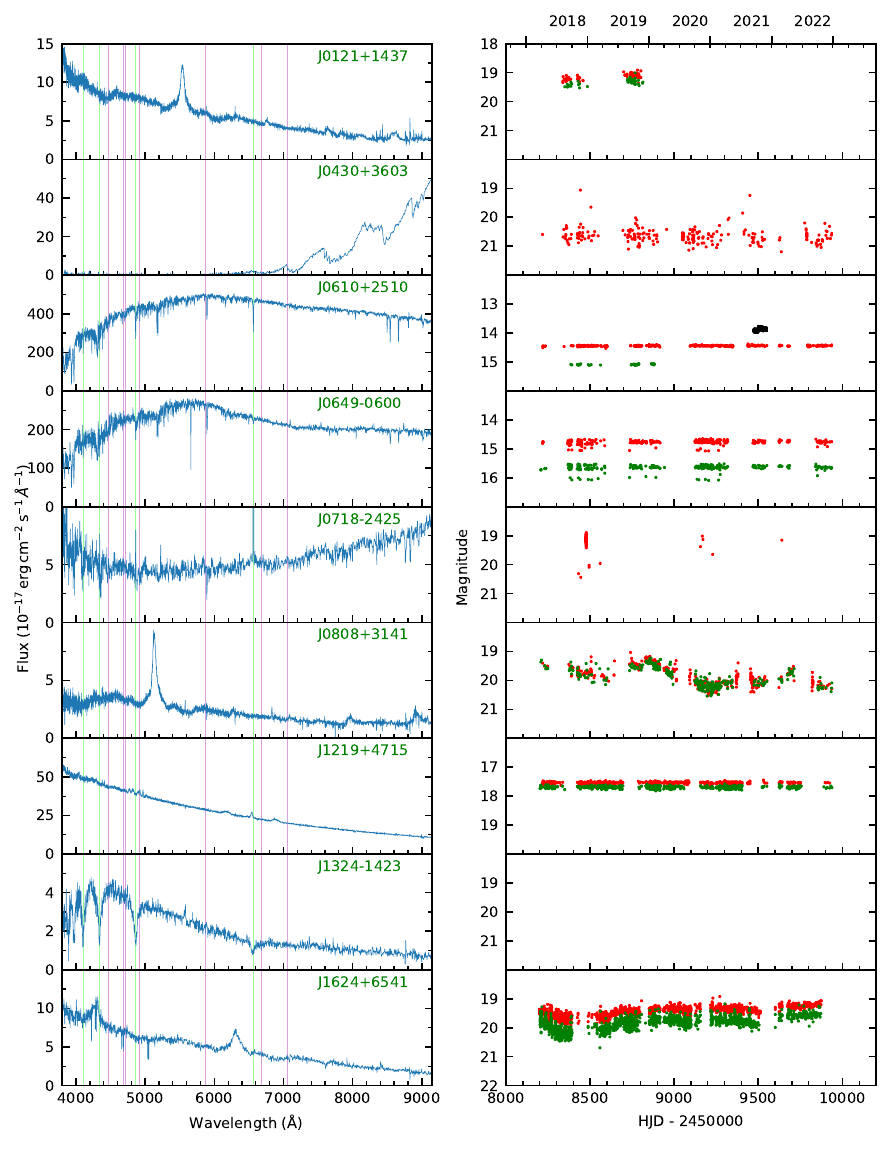}
\caption{\label{fig:NotsCVSpecLC01} \protect\SDSSV\  Spectra (left) and ($r$- and $g$-band)  ZTF  light curves   (right) of systems misclassified in the literature as CVs and CV candidates.}
\end{figure*}

\begin{figure*} 
\includegraphics[width=\textwidth]{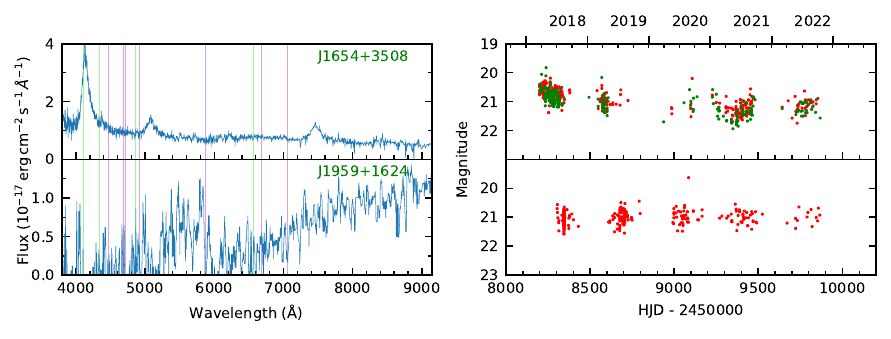}
\caption{\label{fig:NotsCVSpecLC02} Fig.\,\ref{fig:NotsCVSpecLC01} continued.}     
\end{figure*}

Analysis of the \SDSSV\  spectra, ancillary data, and the literature demonstrates that \Nmiscat systems were previously misclassified as CVs or CV candidates. Their \SDSSV\  spectra and ZTF light curves are shown in Figs. \ref{fig:NotsCVSpecLC01} and  \ref{fig:NotsCVSpecLC02}, and we briefly discuss their most likely nature below.

\subsubsection{J0121+1437}
This system was detected as a transient by CRTS (MLS\,160708:012156+143737). The spectrum shows a broad emission line typical of a quasar and is likely to be Ly$\alpha$ at $z\simeq3.5$.  

\subsubsection{J0430+3603}
This object was identified as a transient by MASTER (OT\,J043017.84+360326.9, \citealt{2014ATel.5732....1S}) and subsequently categorised as a U\,Gem dwarf nova by VSX \citealp{2017yCat....102027W}. The \SDSSV\  spectrum reveals a very red slope with strong TiO absorption bands, characteristic of an M-dwarf. The available broad-band photometry is very red and J0430+3603 is located above the location of single low-mass main-sequence stars in the HR diagram. J0430+3603 has a \textit{Spitzer} detection which was categorised as a young stellar object (YSO) in \citet{2014ApJ...786...37B}, which appears consistent with its location in the HR diagram. This object was in the footprint of \textit{GALEX} but was not detected in the ultraviolet. We conclude that J0430+3603 is a single star, and the transient detected by MASTER was very likely a flare.

\subsubsection{J0610+2510}
This object was first identified as a short period variable (NSV\,2853) in 1949, 
and it is classified in VSX as a U\,Gem dwarf nova.  \citet{1996IBVS.4366....1H} obtained CCD photometry for two hours and concluded that it was not a CV. \citet{2000ApJS..128..387L} also obtained photometry and concluded that it was a late G-type star~--~which is consistent with its position on the HR diagram. The ZTF and \textit{TESS} light curves do not show outbursts and we concur with \citet{2000ApJS..128..387L} that it is not a CV. The reason why it was classified as a variable star in the first place remains unclear.

\subsubsection{J0649--0600}
This object, BEST-II lra2b\_01098, was found to be variable by \citet{2009A&A...506..569K}, and it is classified in VSX as a $\delta$ Cepheid type star or a CV. The \SDSSV\  spectrum of J0649$-$0600 resembles a G/K-spectral type, and does not contain any emission lines. Within the \textit{Gaia} HR diagram, J0649--0600 is located in the sub-giant branch.  The ZTF light curve exhibits $0.4$\,mag peak-to-peak variability, and an AOV analysis of the ZTF data reveals this system to be an eclipsing binary with a period of 9.8\,d. We conclude that J0649$-$0600 is not a CV.   

\subsubsection{J0718--2425}
This object was identified as a \textit{Gaia} alert (Gaia18ajg, \citealt{2021A&A...652A..76H})  and classified as a U\,Gem type by VSX. The system is relatively red ($G_\mathrm{BP}-G_\mathrm{RP}\simeq2$), is located slightly above the main sequence in the HR diagram, exhibits a strong infrared excess, and the SDSS spectrum displays very narrow emission lines. All the observational evidence  suggests that J0718$-$2425 is not a CV, and that it may be a YSO.

\subsubsection{J0808+3141}
This object was identified as a transient by MASTER (OT\,J080826.00+314125.6, \citealt{2015ATel.8426....1B}) and subsequently categorised as a CV by VSX. The spectrum shows a broad emission line typical of a quasar and is likely to be Ly$\alpha$ at $z\simeq3.2$.  

\subsubsection{J1219+4715}
This object was classified as a CV candidate based on earlier SDSS spectroscopy   \citep{2006AJ....131..973S} because of its weak H$\alpha$ emission and blue continuum,  characteristics of a CV in outburst or a novalike. However, \citet{2020MNRAS.499.2564G} subsequently classified J1219+4715 as a single magnetic white dwarf with Zeeman-split Balmer emission lines, similar to the enigmatic white dwarf GD356 \citep{1985ApJ...289..732G}.

\subsubsection{J1324--1423}
This object was identified by a transient in  \textit{Gaia} alerts as Gaia17aoi, and has been categorised by VSX as a U\,Gem. \citet{2020MNRAS.494.3799P} raised concerns about the astrometry. The spectrum resembles that of a white dwarf, with no emission lines,  which is consistent with its position in the HR diagram. The SED shows no evidence of a red component and we conclude that this is not a CV but a single white dwarf, and that the nature of the transient detected by \textit{Gaia} remains elusive.

\subsubsection{J1624+6541}
\textit{Gaia} identified this system as a transient, Gaia16aat, which was subsequently classified as a CV by VSX. The spectrum shows a broad emission line typical of a quasar, which is likely to be Ly$\alpha$ at $z\simeq4.2$.  

\subsubsection{J1654+3508}
This was identified as  a transient by CRTS (MLS\,160613:165449+350804); no follow-up studies are reported in the literature. The spectrum shows a broad emission line typical of a quasar, which is likely to be Ly$\alpha$ at $z\simeq2.4$.  We note that this quasar has a large positive parallax, albeit with a large uncertainty.

\subsubsection{J1959+1624}
This object was identified as a \textit{Gaia} alert (Gaia17bqf) and classified as a U\,Gem dwarf nova by VSX. All other \textit{Gaia} observations were non-detections. It is located near the region of the main-sequence occupied by M0 types in the \textit{Gaia} HR diagram. The \SDSSV\  spectrum is very noisy, and shows no indication of emission lines. The available broad-band photometry is consistent with a single cool stellar object without any ultraviolet excess. We conclude that J1654+3508 is not a CV, and that the brightening detected by \textit{Gaia} was most likely due to a stellar flare.

\begin{table*}
\caption{Period bouncers (with candidates shown in italics)  from \allsdss\, and \SDSSV. This sample was used (excluding those with bad astrometry) to characterise the location of period bouncers in the HR diagram (Fig.\,\ref{fig:HRbouncers}). Where the donor mass is not known the  ratio of donor mass to white dwarf mass ($q$) is provided~--~reasonable values for the white dwarf mass imply donor masses consistent with being a period bouncer.}\label{tab:periodbouncers}
\begin{tabular}{llrrclcl}
\hline
SDSS name & Other name & \textit{Gaia} & Distance & Bad astrometry & \Porb & Donor mass & Reference \\ 
 & & $G$ & (pc) & & (h) & (\Msun) & \\
\hline
\textit{J0006+0121} &  & 19.64 & 352 &  & 1.52 &  & This work\\ 
\textit{J0233+0050} & HP\,Cet & 20.16 & 683 & * & 1.6 &  & This work\\ 
\textit{J0845+0339} & V498\,Hya & 20.83 & 1635 & * & 1.45 &  & This work\\ 
\textit{J1434+3340} &  & 20.14 & 363 &  &  &  & This work\\ 
\textit{J0058-0107} & ASASSN-16iw &  &  & * & 1.56 & $q=0.079$ & \citet{2017PASJ...69...75K}\\ 
J0804+5103 & EZ\,Lyn & 17.79 & 143 &  & 1.43 & 0.042 & \citet{2021ApJ...918...58A}\\ 
J0843+2751 & EG\,Cnc & 18.77 & 187 &  & 1.44 & 0.02 & \citet{1998PASP..110.1290P}\\ 
\textit{J0904+4402} & FV\,Lyn & 19.42 & 335 &  & 1.67 &  & \citet{2023MNRAS.524.4867I}\\ 
J1035+0551 &  & 18.78 & 196 &  & 1.37 & 0.06 & \citet{2006Sci...314.1578L}\\ 
J1057+2759 &  & 19.54 & 355 &  & 1.51 & 0.044 & \citet{2017MNRAS.467.1024M}\\ 
\textit{J1212+0136} & V379\,Vir & 17.98 & 152 &  & 1.47 &  & \citet{2008ApJ...674..421F}\\ 
J1216+0520 &  & 20.11 & 377 &  & 1.65 & $<0.04$ & \citet{2006MNRAS.373..687S}\\ 
J1255+2642 & MT\,Com & 19.21 & 347 &  & 1.99 & $<0.05$ & \citet{2005PASP..117..427P}\\ 
J1433+1011 &  & 18.55 & 232 &  & 1.3 & 0.0571 & \citet{2011MNRAS.415.2025S}\\ 
J1435+2336 &  & 18.62 & 209 &  & 1.3 &  & \citet{2022MNRAS.510.6110P}\\ 
J1507+5230 & OV\,Boo & 18.15 & 212 &  & 1.11 & 0.056 & \citet{2007MNRAS.381..827L}\\ 
\textit{J2131$-$0039} & QZ\,Aqr &  &  & * & 1.67 &  & \citet{2023MNRAS.524.4867I}\\ 
J2304+0625 & KX\,Psc & 20.91 &  & * & 1.62 & $q=0.053$ & \citet{2014PASJ...66..116N}\\ \hline

\end{tabular}
\end{table*}
\section{Discussion}\label{section:discussion}

\subsection{Period Bouncers}

The standard evolutionary model of CVs (see \citealt{2023MNRAS.524.4867I} for a recent summary) predicts  that nuclear burning in the donor star stops once mass transfer has eroded its mass to $\simeq0.06\Msun$, at which point the star's structure becomes dominated by electron degeneracy, and it becomes a brown dwarf. As a consequence of the now inverted mass-radius relation, the orbital separations and periods increase as they continue their evolution. CVs that have passed the period minimum are colloquially referred to as ``period bouncers''. QZ\,Lib is an excellent example of a period bouncer, having an orbital period of 92.7\,min, but a very cool ($\lesssim1700$\,K) donor star \citep{2018MNRAS.481.2523P}. The evolution of CVs near the period minimum has been one of the areas where theory and observations show the largest discrepancies: 

(1) The predicted location of the period minimum fell short of the observed value by about ten per cent \citep{1999MNRAS.309.1034K}. Assuming some amount of angular momentum loss in addition to gravitational wave radiation removes that discrepancy \citep{1998PASP..110.1132P,2011ApJS..194...28K}, and this is the currently favoured solution. 

(2) As the evolution of CVs not only slows down towards the period minimum, but also subsequently reverses towards longer periods, an accumulation of systems near the period minimum, the so-called ``period minimum spike'' has been a firm prediction of all CV population models. Yet observed CV samples consistently fail to exhibit this feature \citep[e.g.][]{2002MNRAS.335..513K}, and it was only the SDSS CV sample that confirmed the existence of the period minimum spike \citep{2009MNRAS.397.2170G}. The key conclusion was that the earlier CV samples were too shallow to capture a sufficiently large number of the faint population of period-minimum CVs.

(3) Evolutionary models predict that a large proportion of CVs are period bouncers: \citet{1993A&A...271..149K} suggested that 70\,per\,cent of all CVs are period bouncers whilst \citet{2015ApJ...809...80G} predicted 38$-$60\,per\,cent and Table\,1 in \citet{2018MNRAS.478.5626B} suggests 75\,per\,cent. Despite much observational effort, the roster of known period bouncers remains small:   \citet{2011MNRAS.411.2695P} discussed 22 candidates among the known CVs; \citet{2020MNRAS.494.3799P} estimated a period bouncer fraction of $7-14$~per cent within the 150\,pc CV sample, and \citet{2023MNRAS.524.4867I} concluded that only 0.7\,per cent of all the CVs observed by \allsdss were period bouncers. \citet{2018MNRAS.473.3241H}  used the SDSS Stripe~82 and Palomar Transient Factory photometry for an unbiased search of eclipsing period bouncers, and derived an upper limit on their space density of $\lesssim2\times10^{-5}\,\mathrm{pc^{-3}}$. 

The dedicated spectroscopic follow-up of CV candidates carried out by \SDSSV\  obtained nearly-complete spectroscopy of period-bounce candidate systems, which we use below to derive a more stringent estimate of the space density of period bouncers.

\begin{figure} 
\includegraphics[width=\columnwidth]{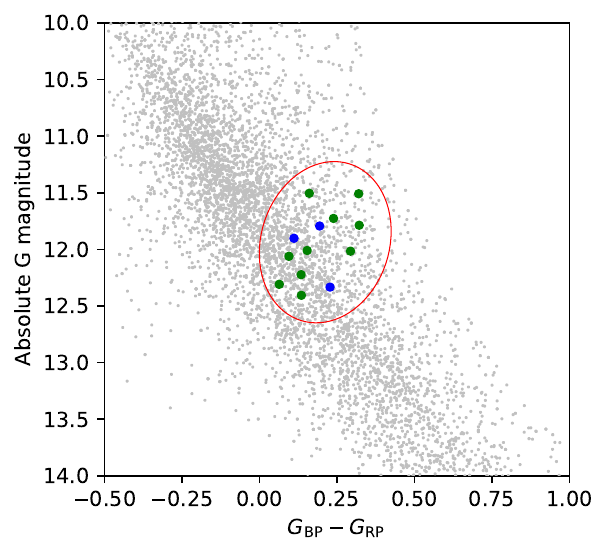}
\caption{\label{fig:HRbouncers} Hertzsprung-Russell (HR) diagram of period bouncers (green dots) and candidates (blue dots) with reliable parallaxes ($\Delta \varpi < 0.2 \times \varpi$) taken from \citet{2023MNRAS.524.4867I} and this paper (see Table \ref{tab:periodbouncers}). The red contour shows the minimal enclosing ellipse (centred on $G_\mathrm{BP}-G_\mathrm{RP}=0.193$, $G_\mathrm{abs}=11.921$ with semi major and minor axes of 0.574 and 0.1159 and rotated by $96.70\degree$) containing the known period bouncers, allowing for $1\,\sigma$ uncertainties in their \textit{Gaia} parameters. The grey dots are  the targets of the \texttt{mwm\_wd} carton within the footprint of the 236 \protect\SDSSV\  plates analysed here.
}
\end{figure}

\begin{figure} 
\includegraphics[width=\columnwidth]{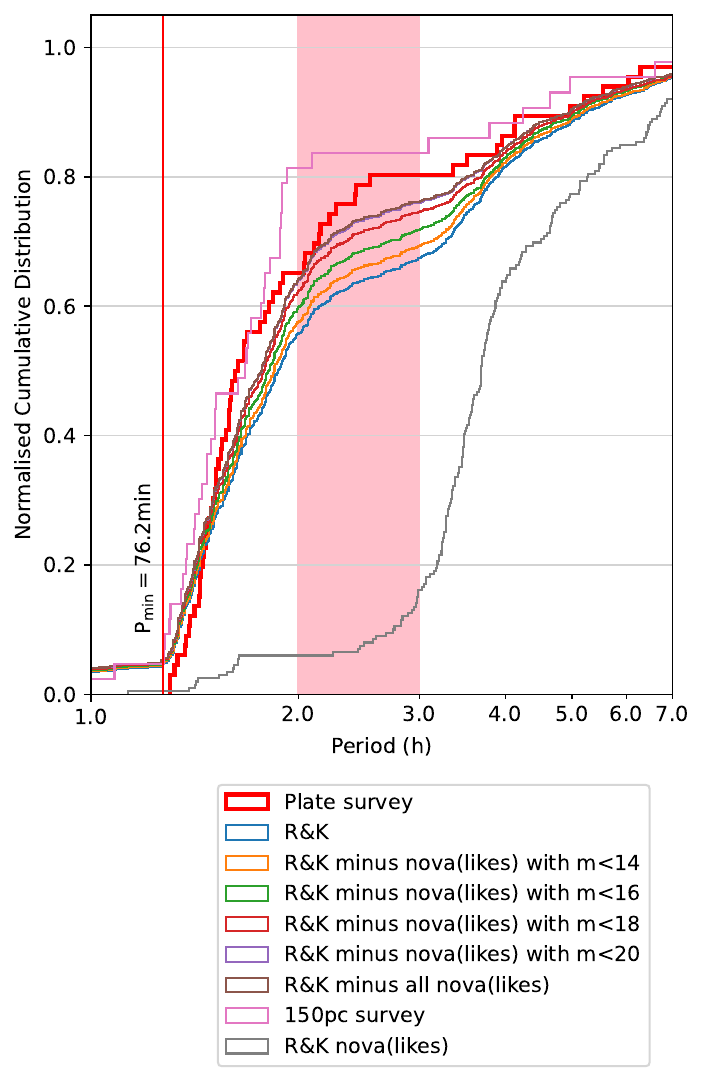}
\caption{\label{fig:PeriodDist}Cumulative distribution of the orbital periods of the CVs observed by \protect\SDSSV, compared to those of the \citet{2003A&A...404..301R} catalogue of CVs and of the 150\,pc volume-limited survey \citep{2020MNRAS.494.3799P}. The period minimum (red line) and ``period gap'' (pink rectangle) are indicated (see text for details). We also include the period distribution of \citet{2003A&A...404..301R} with classical novae and novalikes removed, for a range of limiting magnitudes, as well as the period distribution of classical novae and novalikes alone,  demonstrating the selection effect of these intrinsically bright objects}.    
\end{figure}

\begin{figure*} 
\includegraphics[width=\textwidth]{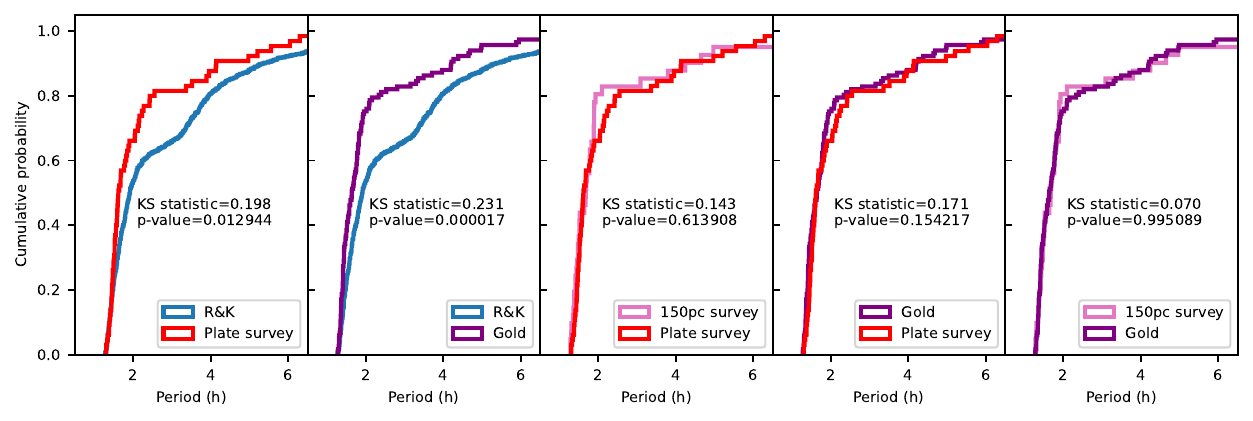}
\caption{\label{fig:kstest} We used the two-sample Kolmogorov–Smirnov (KS) test \citep{1958ArM.....3..469H} to compare pair-wise the cumulative orbital period distributions of the CVs observed by \protect\SDSSV\ (Plate survey), the CVs in the \citet{2003A&A...404..301R} catalogue  (R\&K), the CVs in the volume-limited 150\,pc sample of \citealt{2020MNRAS.494.3799P} (150\,pc) and the CVs in the volume-limited 300\,pc ``Gold'' sample from \citealt{2021MNRAS.504.2420I} (Gold). The KS tests demonstrate that the underlying period distributions of both the Gold sample and the  \protect\SDSSV\ CVs differ from that of  R\&K  at a highly significant level. In contrast, there is no evidence that the period distributions underlying the \protect\SDSSV, 150\,pc and 300\,pc samples differ.}
\end{figure*}

We collected a sample of confirmed and high-confidence period bouncers (Table\,\ref{tab:periodbouncers}) that have SDSS spectroscopy, either obtained as part of \allsdss\ \citep{2023MNRAS.524.4867I} or \SDSSV\ (this paper), and used their \textit{Gaia} astrometry and photometry to identify their location within the HR diagram. We found that period bouncers are closely clustered near the cooling sequence of single white dwarfs  (Fig.\,\ref{fig:HRbouncers}), which is unsurprising as their optical spectra are dominated by the flux from the white dwarf, with only small contributions from the accretion discs and negligible amounts of light from the brown dwarf donors. A number of the systems in Table\,\ref{tab:periodbouncers} stood out as outliers from this clustering. Upon closer inspection, we found that these outliers have poor \textit{Gaia} astrometry, and consequently, we removed all systems with $\Delta\varpi/\varpi\ge0.2$ from the definition of the location of period bouncers in the HR diagram. We then computed the minimal enclosing ellipse \citep{10.1007/BFb0038202} that contains the known period bouncers, where we accounted for the observed uncertainties by displacing each system within its $1\,\sigma$ errors.  We then assume that hitherto undiscovered period bouncers have similar absolute magnitudes and colours to the ones that are known already. We believe that this is a reasonable assumption as the white dwarf temperature is set by the accretion rate \citep{2009ApJ...693.1007T}, and there is so far  no known CV with a quiescent white dwarf temperature $\lesssim9500$\,K \citep{2015AcPPP...2...35S}. Theoretical models predict that the accretion rate should rapidly decrease as systems evolve back from the period minimum, and hence longer period period bouncers should have significantly cooler white dwarfs. However no such system is known, which means that if they exist, they are undetectable by photometric variability (including outbursts or eclipses), X-ray emission, or emission lines, as plenty of data is available to find at least a handful of these hypothetical systems.

Next, we analysed the number of targets within the \texttt{mwm\_wd} carton that fall within this ellipse, and the number of spectroscopic observations of these targets carried out by \SDSSV. We made use of only the \texttt{mwm\_wd} carton as it provides good coverage of the HR diagram location occupied by period bouncers, has the cleanest selection rules, homogeneously covers the entire \SDSSV\  footprint, and has a well-defined ($G\le20$\,mag) magnitude limit. We found 1376 \texttt{mwm\_wd} targets within the ellipse defined by the known period bouncers, of which 1132 were observed during the \SDSSV\  plate program, i.e. the correction factor to account for the incompleteness of the spectroscopic observations is $C_\mathrm{orr}=1.22$. Amongst these 1132 observed targets eight displayed emission lines and these were classified using the methodology described in  section \ref{section:identification}. From these eight we identified only one candidate period bouncer (based on the combination of the signature of the white dwarf being visible in the spectrum, the donor not being visible in the spectrum, the SED being consistent with a  brown dwarf donor and the orbital period significantly greater than the period minimum): J0006+0121 with a period of 91\,min ($N_{\mathrm{obs}}=1$)\footnote{The plate program obtained spectroscopy of three additional 
 candidate period bouncers, J0233+0050 ($\Porb=96$\,min), J0845+0339 with a period of 87\,min and J1434+3340 (no \Porb), which are however all too faint to be included in the \texttt{mwm\_wd} carton.}

We estimated the space density of period bouncers using the technique described in Section\,6.4 of \citet{2023MNRAS.524.4867I}. We first concluded from inspection of Fig.\,\ref{fig:HRbouncers} that all period bouncers would have an absolute magnitude $G_\mathrm{abs}\le12.25$, yielding a limiting distance of $R_\mathrm{lim}=355$\,pc. Following \citet{2023MNRAS.524.4867I} we assumed a scale height of 450\,pc as period bouncers are an old population. Next, we determined the \SDSSV\  sky coverage from the \Nplates plates on which the 1132 \texttt{mwm\_wd} targets were observed. HEALpixs were then used to find the sky coverage of these \Nplates plates by accounting for overlaps and hence the effective volume of the magnitude limited \texttt{mwm\_wd} sample contained within the ellipse defined by the known period bouncers.

Taken together, these assumptions yielded a space density  $\rho_0 \simeq 0.2 \times 10^{-6}\,\mathrm{pc}^{-3}$ for period bouncers. Estimates for the composite space density of all sub-types of CVs range from $\rho_0 = 4.8 \times 10^{-6}\,\mathrm{pc}^{-3}$ \citep{2020MNRAS.494.3799P} to $\rho_0 = 7.8 \times 10^{-6}\,\mathrm{pc}^{-3}$ \citep{2023MNRAS.524.4867I} implying that accreting period bouncers only account for about three\,per\,cent of CVs.

From the above analysis, we can only conclude that the large number of period bouncer predicted by CV models  either do not exist, or that the majority of them look so distinctly different from the small sample of known period bouncers that they escape spectroscopic (our analysis) and photometric \citep{2018MNRAS.473.3241H} detections. Our conclusions echo those of \citet{1998PASP..110.1132P}, \textit{``It is probably necessary to have some means of destroying CVs before they reach the predicted very high space densities.''} Possible scenarios to ``destroy'' CVs include either a merger, maybe as a result of a final classical nova eruption (e.g. via consequential angular momentum loss, \citealt{2016MNRAS.455L..16S,2016ApJ...817...69N}), or the secondary contracting within its Roche lobe, thereby terminating mass transfer. The latter option was suggested by \citet{1998PASP..110.1132P}, arguing that the final state of CVs may be white dwarfs with detached, planet-like companions~--~and the ultimate test of that hypothesis would be a more sensitive survey for apparently single white dwarfs eclipsed by their (at optical wavelengths) invisible ultra-low-mass companions. 

\subsection{Orbital period distribution}
We have obtained periods for  \Nperiods of the \NCVs CVs in the \SDSSV\  plate survey.  The main reasons why the remaining systems have no period measurements are due to either a low inclination (which will reduce the amplitude of any orbital modulation), faintness (i.e. low signal-to-noise ratio of the time-series photometry or spectroscopy), or sparse sampling of the available time-series data. With the possible exception of low \mdot\ short-period CVs (which do not have much orbital modulation from the hot spot and are inherently fainter) these factors are not likely to correlate with the period. 

The orbital period distribution of the \SDSSV\  CVs is shown in Fig.\,\ref{fig:PeriodDist} where it is compared to that of the \citet{2003A&A...404..301R} sample, which is a very heterogeneous collection of CVs, and the  150\,pc volume-limited sample of \citep{2020MNRAS.494.3799P}. The period gap, which has been a defining feature of the observed period distribution of CVs and motivated the widely accepted idea of ``disrupted magnetic braking'' \citep{1983ApJ...275..713R,2001ApJ...550..897H,2010A&A...513L...7S}, is at best marginally detected in the \SDSSV\  sample which is in clear contrast with the period distribution of the CVs from the Ritter and Kolb (R\&K) catalogue. Classical novae (post-eruption) and novalike variables are inherently bright (Figure\,15 in \citealt{2023MNRAS.524.4867I}) and are over-represented in \citet{2003A&A...404..301R}. They also typically have orbital periods above $\simeq3$\,h. We therefore removed classical novae and novalikes from \citet{2003A&A...404..301R} and the resulting period distribution (see Fig.\,\ref{fig:PeriodDist}) is then closely aligned with that of the \SDSSV\ sample. To illustrate the effect of different limiting magnitudes, we also include period distributions of the R\&K sample cut at $m<14, 16, 18$, and 20, which clearly shows that different limiting magnitudes has the largest effect in the $\simeq3-4$\,h orbital period range. 

In order to assess the statistical significance of the period distributions of different observed CV populations, we carried out the two-sample Kolmogorov–Smirnov (KS) test \citep{1958ArM.....3..469H} pairwise between two volume-limited samples, the 150\,pc sample of \citet{2020MNRAS.494.3799P} and the 300\,pc ``Gold'' sample of \citep{2021MNRAS.504.2420I} and two magnitude-limited samples, the \SDSSV\ CVs analysed in this paper and those in the \citet{2003A&A...404..301R} catalogue (R\&K). We truncated the period distributions for $\Porb<75$\,min to remove the AM\,CVn systems, as we are interested in comparing the properties of ``normal'' CVs with hydrogen-rich donor stars. The KS test evaluates the likelihood of the samples being compared sharing the same underlying orbital period distribution. The KS statistic measures the maximum difference in the cumulative probability distributions and we used the \textsc{scipy.stats.ks\_2samp} function  to calculate a p-value. For instance, a p-value of less than $0.05$ implies that the surveys are drawn from different distributions with $95$ per cent confidence. 

The main difference between the 150\,pc and 300\,pc samples is that the former has a substantially higher completeness  ($77\pm10$~per cent) than the latter ($\simeq50$~per cent). However the 150\,pc sample is much smaller (42 systems) than the 300\,pc sample  (151 systems). The cumulative distributions of the two samples are visually very similar (right-most panel in Fig.\,\ref{fig:kstest}), and the KS indicates a very low probability of the two being drawn from different underlying populations. We hence conclude that despite the lower completeness, the 300\,pc sample is  representative of the intrinsic CV population. 

Comparing the period distribution of the \SDSSV\ CVs to those of the 150\,pc and 300\,pc samples (third and second panels from the right in Fig.\,\ref{fig:kstest}, respectively) reveals a shortfall of systems around $P\simeq2$\,h, however, the p-values determined from the KS tests are sufficiently large that there is no statistical evidence that the period distribution of the \SDSSV\ CV sample differs from those of the volume-complete ones.

However, the 300\,pc CVs and the R\&K sample (second panel from the left in Fig.\,\ref{fig:kstest}) have very clearly distinct period distributions: the R\&K sample shows distinct changes in the slope at $\simeq2$\,h and $\simeq3$\,h, which \citet{2006MNRAS.373..484K} interpreted as clear signatures of the lower and upper edge of the period gap, respectively. In contrast to this, the 300\,pc CV sample shows a steep break in the slope of the cumulative distribution at $\simeq2$\,h, but much less-pronounced structure at longer orbital periods~--~there is a small hint of a break in the slope at $\simeq4$\,h. Based on the KS test, the two samples have intrinsically different period distributions at a very high significance ($>99.9$~per cent).

Similarly, the \SDSSV\ CVs and the R\&K sample (left-most panel in Fig.\,\ref{fig:kstest}) differ distinctively, with the number of short-period \SDSSV\ CVs increasing well beyond the canonical $\simeq2$\,h lower edge of the period gap, and not exhibiting a strong break at $\simeq3$\,h. Based on the KS test, the two samples have intrinsically different period distributions at a very high significance ($\simeq98$~per cent). Whereas we stress that the \SDSSV\ sample is neither complete nor subject to a homogeneous selection of the CV candidates that were targeted for spectroscopy, it approximates the inherent period distribution revealed by the volume-limited samples better than the R\&K sample~--~most likely because of the large limiting magnitude of SDSS.

We conclude that the observational evidence for the period gap is primarily based on the very heterogeneous R\&K sample, and is linked to the large fraction of classical novae and novalike variables within that sample. If the period gap is indeed an intrinsic feature of the CV population, and a signature of disrupted magnetic braking, its early identification \citep{1980MNRAS.190..801W} would be based on a ``lucky'' selection effect among the earliest CV discoveries.

We note in passing that the \SDSSV\  period distribution, as well as those of the 300\,pc sample, show a slight flattening between $\simeq3$\,h and $\simeq5$\,h which is consistent with the predictions of \citet{2021NatAs...5..648S} for CVs where the formation of magnetic white dwarfs undergoes a detached phase within that period range, resulting in a decrease in the number of accreting CVs. However, we stress that a larger volume-limited CV sample is required before this feature would be sufficiently secure to provide robust support to the model of \citet{2021NatAs...5..648S}.

Finally, it is worth noting that the orbital periods of the new CVs found by \SDSSV, where we were able to obtain unambiguous measurements, are all below 94\,min,  which is typical of old, low accretion rate systems. This reinforces the findings of \citet{2009MNRAS.397.2170G} and suggests that \SDSSV\  will continue to increase the proportion of short-period CVs among the known CV population. 

\subsection{CV sub-types}
\begin{figure} 
\includegraphics[width=\columnwidth]{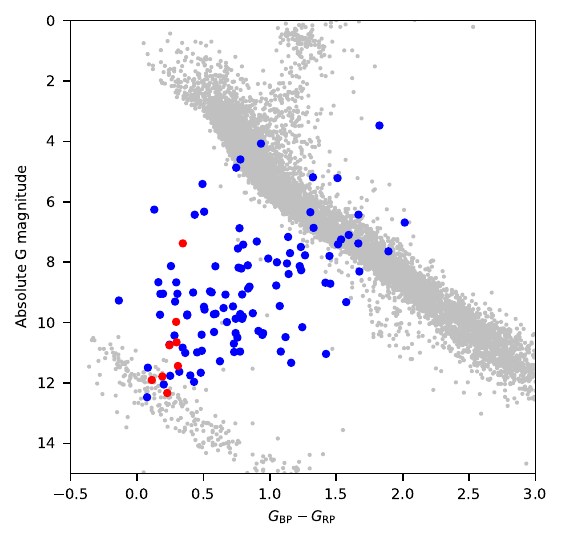}
\caption{\label{fig:HR Diagram} HR diagram showing the \Nprev previously known CVs in blue and the eight new discoveries in red. The grey dots are an astrometrically clean sample of \textit{Gaia} objects chosen to outline the main sequence and the white dwarf cooling sequence}
\end{figure}

The distribution of the new and previously known CVs is given  in Table\,\ref{tab:types}. Taken in conjunction with their distribution in the HR diagram (Fig.\,\ref{fig:HR Diagram}) we notice that the new discoveries include a novalike variable in the expected area (see Fig.\,8 in \citealt{2021MNRAS.504.2420I}). All of the other new discoveries are close to the white dwarf cooling sequence, i.e. the area associated with short-period CVs with low accretion rates,  experiencing infrequent outbursts. This demonstrates the importance of a spectroscopic survey that compensates for the traditional bias of CV searches based on large-amplitude photometric variability.  In passing, we note that there are \Nmags magnetic CVs (14\,per\,cent of the total) in the \SDSSV\  CV sample, which is similar to the fraction found in  the \allsdss sample \citep{2023MNRAS.524.4867I}, but lower than in the volume-limited 150\,pc sample, probably due to the, on average, lower accretion rates in polars when compared to non-magnetic CVs \citep{2002MNRAS.335....1W,2005ApJ...622..589A}.


\begin{table}
\caption{The CV sub-types of the new and previously known systems}\label{tab:types}
\begin{tabular}{llll}\hline
CV type & Previously known & New & Total \\ \hline
U\,Gem  & 26 & 0 & 26 \\ 
ER\,UMa  & 3 & 0 & 3 \\ 
SU\,UMa  & 36 & 1 & 37 \\ 
WZ\,Sge  & 9 & 4 & 13 \\ 
Z\,Cam  & 1 & 0 & 1 \\ 
AM\,CVn  & 3 & 0 & 3 \\ 
Classical nova  & 5 & 0 & 5 \\ 
Novalike  & 1 & 1 & 2 \\ 
Polar - AM Her  & 14 & 1 & 15 \\ 
Intermediate polar - DQ Her  & 2 & 0 & 2 \\ 
Magnetic CV  & 1 & 0 & 1 \\ 
Dwarf nova  & 7 & 0 & 7 \\ 
Unclassified  & 2 & 1 & 3 \\ 
\hline\textbf{Total} & 110 & 8 & 118 \\ \hline 

\end{tabular}
\end{table}

\section{Summary} \label{section:summary}
\SDSSV\  is, for the first time, carrying out a dedicated survey of white dwarfs, both single and in binaries. We have analysed the \SDSSV\  spectra of CVs and CV candidates observed as part of the final plug-plate operations of SDSS, and we discovered eight new CVs. None of the new CVs displays noticeable changes in brightness in their ZTF light curves, underlining the selection effects in existing CV samples that are largely based on photometric variability, and hence the importance of unbiased spectroscopic surveys. We did not discover any new CVs within 150\,pc which is consistent with the relatively high completeness of this volume-limited sample \citep{2020MNRAS.494.3799P}.

\SDSSV\  also observed \Nprev previously known CVs and CV candidates, for \Nspec of these, the SDSS data represents the first spectroscopic confirmation of their CV nature. Vice versa, the \SDSSV\  spectroscopy disproves the CV nature of \Nmiscat systems that were previously classified as CVs or CV candidates. 

We measured \Nnewperiods\ new orbital periods from the analysis of radial velocities and time-series photometry. The period distribution of the \SDSSV\  CV sample does not provide strong evidence for a period gap, which contrasts strongly with the orbital period distribution of the large, but very heterogeneous \citep{2003A&A...404..301R} sample. We suggest that future modelling of the CV population should aim to reproduce the period distribution of well-defined samples. 

Lastly, but most importantly, we demonstrate that it is unlikely that there is a large population of period bounce CVs awaiting discovery, which represents a major challenge to CV evolution theory. Whilst we cannot totally rule out the possibility of  hypothetical period bouncers with a  radically different appearance from the handful of known systems we consider this scenario improbable. The implications are, most likely, that CVs that evolve past the period minimum either get destroyed (e.g. by a merger) or become detached and therefore very difficult to detect. The Legacy Survey of Space and Time survey will provide an opportunity to search for a population of cool white dwarfs that are eclipsed by planet-mass and size objects with periods of $\sim80-120$\,min.

\SDSSV\  will continue to survey both the northern and southern hemisphere for single and binary white dwarfs, and the growing sky coverage will provide increasingly stronger constraints on the intrinsic properties of the Galactic CV population.

\section*{Acknowledgements}

Funding for the Sloan Digital Sky Survey V has been provided by the Alfred P. Sloan Foundation, the Heising-Simons Foundation, and the Participating Institutions. SDSS acknowledges support and resources from the Center for High-Performance Computing at the University of Utah. The SDSS web site is \url{www.sdss5.org}. SDSS is managed by the Astrophysical Research Consortium for the Participating Institutions of the SDSS Collaboration, including the Carnegie Institution for Science, Chilean National Time Allocation Committee (CNTAC) ratified researchers, the Gotham Participation Group, Harvard University, The Johns Hopkins University, L'Ecole polytechnique f\'{e}d\'{e}rale de Lausanne (EPFL), Leibniz-Institut f\"{u}r Astrophysik Potsdam (AIP), Max-Planck-Institut f\"{u}r Astronomie (MPIA Heidelberg), Max-Planck-Institut f\"{u}r Extraterrestrische Physik (MPE), Nanjing University, National Astronomical Observatories of China (NAOC), New Mexico State University, The Ohio State University, Pennsylvania State University, Smithsonian Astrophysical Observatory, Space Telescope Science Institute (STScI), the Stellar Astrophysics Participation Group, Universidad Nacional Aut\'{o}noma de M\'{e}xico, University of Arizona, University of Colorado Boulder, University of Illinois at Urbana-Champaign, University of Toronto, University of Utah, University of Virginia, Yale University, and Yunnan University.

The Liverpool Telescope is operated on the island of La Palma by Liverpool John Moores University in the Spanish Observatorio del Roque de los Muchachos of the Instituto de Astrofisica de Canarias with financial support from the UK Science and Technology Facilities Council.

The authors are honored to be permitted to conduct astronomical research on Iolkam Du’ag (Kitt Peak), a mountain with particular significance to the Tohono O’odham.

Based on observations obtained with the Samuel Oschin 48$-$inch Telescope at the Palomar Observatory as part of the Zwicky Transient Facility project. ZTF is supported by the National Science Foundation under Grant No. AST-1440341 and a collaboration including Caltech, IPAC, the Weizmann Institute for Science, the Oskar Klein Center at Stockholm University, the University of Maryland, the University of Washington, Deutsches Elektronen-Synchrotron and Humboldt University, Los Alamos National Laboratories, the TANGO Consortium of Taiwan, the University of Wisconsin at Milwaukee, and Lawrence Berkeley National Laboratories. Operations are conducted by COO, IPAC, and UW

CRTS is supported by the U.S. National Science Foundation under grants AST-0909182 and CNS-0540369. The work at Caltech was supported in part by the NASA Fermi grant 08-FERMI08-0025, and by the Ajax Foundation. The CSS survey is funded by the National Aeronautics and Space Administration under Grant No. NNG05GF22G issued through the Science Mission Directorate NEOs Observations Program. 

This research has made use of the International Variable Star Index (VSX) data base, operated at AAVSO, Cambridge, Massachusetts, USA. 

This paper includes data collected by the TESS mission. Funding for the TESS mission is provided by the NASA's Science Mission Directorate.

This work has made use of data from the European Space Agency (ESA) mission {\it Gaia} (\url{https://www.cosmos.esa.int/gaia}), processed by the {\it Gaia} Data Processing and Analysis Consortium (DPAC, \url{https://www.cosmos.esa.int/web/gaia/dpac/consortium}). Funding for the DPAC has been provided by national institutions, in particular the institutions participating in the {\it Gaia} Multilateral Agreement.  This research has made use of NASA's Astrophysics Data System. This research has made use of the VizieR catalogue access tool, CDS, Strasbourg, France. 

This project has received funding from the European Research Council (ERC) under the European Union’s Horizon 2020 research and innovation programme (Grant agreement No. 101020057). 

MRS acknowledges support from FONDECYT (grant 1221059) and ANID (Millennium Science Initiative NCN19\_171).

This work was supported by the Deutsche Forschungsgemeinschaft under grant Schw536/37-1.

G.T. was  supported by grants IN109723 and IN110619 from the Programa de Apoyo a Proyectos de Investigación e Innovación Tecnológica (PAPIIT).

OT was supported by a FONDECYT project 321038

\section*{Data Availability}
\SDSSV\  data will be publicly available at the end of the proprietary period. The reduced LT light curves presented here will be shared upon reasonable request to the corresponding author. The other data used in this article are available from the sources referenced in the text. 


\bibliographystyle{mnras}
\bibliography{refs,others} 



\clearpage
\appendix

\onecolumn

\begin{landscape}
\section{Table of CVs observed}

\setlength{\LTcapwidth}{1.2\textwidth}
\begin{longtable}[c]{lllllrlllll}
\caption{CVs from the Plate Survey. Distances are from \citet{2021AJ....161..147B} and variable types from \citet{2006SASS...25...47W} unless indicated in blue. The references for the initial discovery, spectrum and period (where known) are also shown. New discoveries, orbital periods and identifications are shown in blue. Suffixes to the classifications : Tentative values, * Superhump periods, \dag~Low accretion SU\,UMa, \ddag~CVs with unusual state changes, \S~Period bouncers, +E Eclipsing. \label{tab:cv_master_table}}\\

\hline
\multirow{2}{*}{SDSS} & \multirow{2}{*}{Name} & 
\multirow{2}{*}{\begin{tabular}[c]{@{}l@{}}Gaia EDR3\\ source\_id\end{tabular}}  &
\multirow{2}{*}{\begin{tabular}[c]{@{}l@{}}Period \\ (h)\end{tabular}} & 
Gaia $G$ & 
Distance & 
Variable & 
\multirow{2}{*}{Cartons} & \multicolumn{3}{c}{References}                         \\ \cline{9-11} 

 &    &     &   & (mag) & (pc) & Type  &    & \multicolumn{1}{l}{Disc} & \multicolumn{1}{l}{Sp} & P \\ \hline
 \hline
 \endfirsthead

\hline
\multirow{2}{*}{SDSS} & \multirow{2}{*}{Name} & 
\multirow{2}{*}{\begin{tabular}[c]{@{}l@{}}Gaia EDR3\\ source\_id\end{tabular}}  &
\multirow{2}{*}{\begin{tabular}[c]{@{}l@{}}Period \\ (h)\end{tabular}} & 
Gaia $G$ & 
Distance & 
Variable & 
\multirow{2}{*}{Cartons} & \multicolumn{3}{c}{References}                         \\ \cline{9-11} 

 &    &     &   & (mag) & (pc) & Type  &    & \multicolumn{1}{l}{Disc} & \multicolumn{1}{l}{Sp} & P \\ \hline
 \hline
 \endhead
 \hline
 \endfoot
 \hline
 \endlastfoot
\textcolor{blue}{J000600.15+012129.8} &  & 2738755406045571968 & \textcolor{blue}{1.52(10)} & 19.57 & 352 & \textcolor{blue}{WZ Sge:\S :} & gg,u2,wd & 1 & 1 & 1 \\ 
J003827.05+250925.0 & 1RXS J003828.7+250920 & 2806802123399581056 & 2.26826(2) & 18.63 & 503 & SU UMa & cv,gg,u1,u2 & 28 & 12 & 84 \\ 
J003941.08+005427.5 & SDSS J003941.06+005427.5 & 2543387617312121216 & 1.523(2) & 20.76 & 1078 & WZ Sge: & cv,u1,u2,u4 & 53 & 53 & 74 \\ 
J004335.16-003729.7 & SDSS J004335.14-003729.8 & 2530961280492678528 & 1.3721(14) & 19.85 & 447 & \textcolor{blue}{WZ Sge:} & cv,gg,u2,wd & 115 & 115 & 33 \\ 
J014227.07+001729.8 &  & 2510205490257050496 & \textcolor{blue}{1.88(4)} & 19.89 & 697 & Polar: & cv,u1,u2 & 105 & 105 & 1 \\ 
J015543.47+002807.4 & FL Cet & 2507796391561705728 & 1.452389(1) & 18.65 & 317 & Polar+E & cv,gg,u1,u2 & 54 & 54 & 119 \\ 
J020712.72-014116.3 & SDSSJ020712.71-014116.2 & 2494386992562185088 & \textcolor{blue}{1.55(15)} & 20.27 &  & Polar: & cv & 8 & 8 & 1 \\ 
J020809.57+565239.7 &  & 505110900157013376 &  & 18.77 & 2961 & \textcolor{blue}{CV} & u4 & 39 & 1 &  \\
J021008.33+571121.0 & UV Per & 457106501671769472 & 1.557(3) & 17.75 & 250 & SU UMa & cv & 16 & 16 & 2 \\ 
J021315.48+533822.9 & MASTER OT J021315.37+533822.7 & 455380951309048320 & 2.549(7)* & 20.05 & 3224 & SU UMa & cv & 63 & 1 & 71 \\ 
J022102.79+732245.0 & ASASSN-14jq & 546910213373341184 & 1.32427(31)* & 19.94 & 425 & WZ Sge & cv,gg,wd & 24 & 1 & 121 \\ 
J022623.34+711831.5 & AM Cas & 545338083544363392 & 3.9576(3) & 14.37 & 420 & U Gem & cv,gg,u1,u2 &  &  &  \\
J023322.61+005059.4 & HP Cet & 2500552912036565120 & 1.601(2) & 20.12 & 683 & WZ Sge:\S : & cv & 54 & 94 & 94 \\ 
J024131.07+593630.5 & ASASSN-16jv & 464373929923792384 &  & 19.84 & 607 & U Gem & cv & 24 & 1 &  \\
J040834.99+511448.2 & FO Per & 250700853140894976 & 4.126(12) & 16.15 & 587 & U Gem & cv & 17 & 97 & 87 \\ 
J041844.44+510731.2 & NS Per & 271562333932151040 & \textcolor{blue}{6.296(1)} & 17.43 & 1076 & U Gem & cv & 77 & 97 & 1 \\ 
J042609.34+354144.5 & MASTER OT J042609.34+354144.8 & 176429285061830144 & 1.5720(7) & 16.14 & 184 & SU UMa & cv,gg,u1,u2 & 37 & 75 & 41 \\ 
J043127.01+352713.0 & ASASSN-16og & 174868189003974016 &  & 20.37 & 1677 & SU UMa & cv & 24 & 1 &  \\
J044321.37+472125.8 & V0392 Per & 254361745823908736 & 1.5841(4) & 16.17 & 3453 & Nova & cv & 21 &  & 61 \\ 
J050503.07+222529.8 & ASASSN-15su & 3418268607216934016 & 1.608(7)* & 19.12 & 588 & SU UMa & cv & 24 & 1 & 99 \\ 
J052433.47-070527.9 & CSS 101108:052433-070528 & 3207332631484540416 &  & 20.59 & 1538 & SU UMa: & cv & 36 & 1 &  \\
J052834.08+331821.7 & QZ Aur & 3449050362952844288 & 8.579910(1) & 16.87 & 2528 & Nova & cv &  &  & 42 \\ 
J052859.54-033352.4 & V1159 Ori & 3210749947983249408 & 1.49272(3) & 14.28 & 351 & ER UMa & cv & 51 & 51 & 69 \\ 
J061204.48+252832.6 & HQ Gem & 3426743917723751680 &  & 19.31 & 3838 & U Gem & cv &  & 1 &  \\
J062157.13+404236.1 & ASASSN-18bz & 956899025419314816 &  & 19.9 & 3324 & U Gem & cv & 79 & 1 &  \\
\textcolor{blue}{J062406.13+400743.8} &  & 956670430081266176 &  & 18.54 & 1728 & \textcolor{blue}{NL} & u1,u2 & 1 & 1 &  \\
J062429.71+002105.8 & ASASSN-15cu & 3120280997380365824 & \textcolor{blue}{2.429755(4)} & 18.86 & 468 & \textcolor{blue}{SU UMa} & cv & 40 & 1 & 1 \\ 
J062601.53-035523.0 & Gaia17cdo & 3116642228070616192 &  & 19.19 & 1184 & DN & cv & 111 & 1 &  \\
J062724.39+161329.6 & Gaia18atj & 3369335048381292032 & 10.428 & 18.15 & 3919 & U Gem & cv &  &  &  \\
J062953.89-033508.4 & 1RXS J062954.6-033520 & 3105034928633782400 & 9.12 & 13.53 & 652 & Z Cam & cv & 64 & 1 & 108 \\ 
J063047.05+023931.2 & ASASSN-16ow & 3124075927403447936 & 2.1425(13)* & 20.57 & 3374 & SU UMa & cv & 47 & 1 & 71 \\ 
\textcolor{blue}{J063301.11+030334.1} &  & 3130088774241620608 &  & 18.3 & 242 & \textcolor{blue}{WZ Sge:} & wd & 1 & 1 &  \\
J063304.77+032422.8 & Gaia17ber & 3130133544980839680 &  & 19.1 & 2548 & \textcolor{blue}{U Gem} & cv & 100 & 1 &  \\
\textcolor{blue}{J063516.63+030313.6} &  & 3130064176963184896 & \textcolor{blue}{1.4717(2)} & 18.95 & 643 & \textcolor{blue}{Polar:} & wd & 1 & 1 & 1 \\ 
J063844.17+181611.4 & UV Gem & 3371136087085232128 & 2.226(5)* & 18.81 & 722 & SU UMa & cv &  & 1 & 96 \\ 
\textcolor{blue}{J070533.79-125004.7} &  & 3044836838812715264 &  & 18.7 & 246 & \textcolor{blue}{CV:} & wd & 1 & 1 &  \\
J070816.61-124544.8 & ASASSN-16an & 3044908474565329280 &  & 19.16 & 2071 & U Gem & cv & 60 & 1 &  \\
J073339.29+212200.8 & CSS 091111:073339+212201 & 865345406593708160 &  & 19.37 & 1961 & U Gem & cv & 43 & 43 &  \\
J073758.55+205544.6 & SDSS J073758.55+205544.5 & 673118544624392576 &  & 19.79 & 974 & SU UMa: & cv & 13 & 1 &  \\
J073817.74+285519.7 & SDSS J073817.75+285519.7 & 878597549661945984 & \textcolor{blue}{5.555(1)} & 18.46 & 1204 & U Gem+E & cv,gg,u1,u2 & 113 & 113 & 1 \\ 
J074222.55+172806.7 & MLS 110309:074223+172807 & 671077782322909696 &  & 20.04 & 1828 & U Gem & cv & 36 & 114 &  \\
J074419.74+325448.2 & CSS 091029:074420+325448 & 881969129052843520 & 0.55 & 19.83 &  & AM CVn & cv & 36 & 43 & 95 \\ 
J074859.55+312512.5 & SDSS J074859.55+312512.6 & 880821067114616832 & 1.39946616(2) & 18.37 & 542 & SU UMa+E & cv,gg,u1,u2,wd & 13 & 15 & 26 \\ 
J074917.11+365427.9 & SDSSJ074917.11+365428.0 & 919046658301858048 & \textcolor{blue}{1.3401(1)} & 19.17 & 820 & Polar: & cv & 8 & 8 & 8 \\ 
J075107.51+300628.3 & SDSS J075107.50+300628.4 & 878968875354903040 & 1.3918(3)* & 19.68 & 1366 & SU UMa & cv,u1,u2,wd & 13 & 1 & 67 \\ 
J075233.17+294339.5 &  & 875941782403757056 & \textcolor{blue}{1.4455(1)} & 18.5 & 702 & \textcolor{blue}{DN} & gg,u1,u2 & 98 & 1 & 1 \\ 
J075240.45+362823.2 & EU Lyn & 918840426857988736 &  & 20.0 & 631 & Polar+E & cv & 113 & 113 & 48 \\ 
J075331.99+375800.7 & CSS 100407:075332+375801 & 919213474830239616 & \textcolor{blue}{1.5363(1)} & 19.21 & 1699 & MCV: & cv,u1,u2 & 36 & 43 & 1 \\ 
J075540.64+264618.1 & V0516 Gem & 874545471356863104 & \textcolor{blue}{1.4883(1)} & 18.74 & 359 & \textcolor{blue}{Polar:} & cv & 59 & 1 & 1 \\ 
J080538.98+354054.9 & CSS 170417:080539+354055 & 906152376204919040 & \textcolor{blue}{5.2069(1)} & 20.69 & 1770 & U Gem & cv & 38 & 1 & 1 \\ 
J080853.73+355053.6 & CSS 080416:080854+355053 & 906214223733883264 &  & 19.98 & 1807 & U Gem & cv,u2 & 106 & 3 &  \\
J081931.24+213338.0 & MLS 151206:081931+213338 & 676203568093232512 &  & 19.46 & 2768 & \textcolor{blue}{IP:} & cv,u2 & 110 & 86 &  \\
J082253.13+231300.6 & SDSS J082253.12+231300.6 & 678153002209054336 &  & 19.8 & 3077 & \textcolor{blue}{CV:} & cv & 45 & 45 &  \\
J083619.15+212105.2 & CC Cnc & 664787098343076992 & 1.764(12) & 17.45 & 493 & SU UMa & cv,gg,u1,u2 & 7 & 7 & 73 \\ 
J084400.09+023919.2 & V0495 Hya & 3079207606650040320 & \textcolor{blue}{4.9708(6)} & 18.42 & 1467 & U Gem & cv & 113 & 113 & 1 \\ 
J084413.66-012807.9 & CSS 090331:084414-012807 & 3072558340145828864 &  & 20.37 & 2153 & AM CVn & cv,u1,u2 & 43 & 43 &  \\
J084555.06+033929.3 & V0498 Hya & 581352651334960512 & 1.4536(14)* & 20.72 & 1635 & \textcolor{blue}{WZ Sge:\S :} & cv & 96 & 43 & 96 \\ 
J085107.39+030834.5 & CT Hya & 581110831791334400 & 1.59612(3)* & 18.54 & 579 & SU UMa & cv & 17 & 113 & 66 \\ 
J090246.54-014201.9 & ASASSN-15gv & 5764187028557281280 & \textcolor{blue}{2.04(10):} & 20.58 & 1734 & \textcolor{blue}{SU UMa:} & cv & 24 & 1 & 1 \\ 
J090344.25-013326.1 &  & 5764195103095781120 &  & 19.53 & 278 & AM CVn & gg,u2,wd & 85 & 1 &  \\
J090403.48+035501.2 & SDSS J090403.48+035501.2 & 579909164365936512 & 1.4336(1) & 19.31 & 295 & WZ Sge: & cv,gg,u1,u2,wd & 115 & 115 & 115 \\ 
J091410.72+013732.9 & ASASSN-16as & 3843535378146405888 & 6.043(7) & 17.69 & 1016 & U Gem & cv & 46 & 112 & 112 \\ 
J092614.30+010557.4 &  & 3844016380122912640 & \textcolor{blue}{1.4723(1)} & 19.47 & 385 & Polar & gg,u1,u2,wd & 6 & 6 & 1 \\ 
J092620.42+034542.3 & GUVV J092620.4+034541.8 & 3845719592354419712 &  & 19.7 & 933 & \textcolor{blue}{SU UMa} & cv,u1,u2 & 56 & 1 &  \\
J093205.24+034332.5 & ASASSN-17od & 3850925956004914816 &  & 17.75 & 633 & \textcolor{blue}{SU UMa:} & cv & 24 & 1 &  \\
J093238.20+010902.5 & SDSS J093238.21+010902.5 & 3841337699215167232 &  & 19.4 & 3145 & \textcolor{blue}{Polar:} & cv & 113 & 113 &  \\
J104245.16+371819.9 & SDSS J104245.15+371819.8 & 751777469035821312 & \textcolor{blue}{1.82(4)} & 19.84 & 507 & Polar & gg,u2 & 8 & 8 & 1 \\ 
J112332.03+431717.5 & CSS 090602:112332+431717 & 783809506927814912 &  & 19.65 & 1261 & \textcolor{blue}{SU UMa:} & cv,u1,u2 & 124 & 43 &  \\
J113122.38+432238.7 & MR UMa & 772038105376131456 & 1.518(3) & 17.92 & 328 & SU UMa & cv,gg,u1,u2 & 88 & 109 & 35 \\ 
J115419.06+575750.8 & SDSS J115419.06+575750.9 & 845441772229861760 &  & 20.42 & 1144 & ER UMa & cv & 8 & 8 &  \\
J121913.06+204937.6 & SDSS J121913.04+204938.3 & 3952276665815698560 & \textcolor{blue}{1.3018(1)} & 19.19 & 274 & \textcolor{blue}{WZ Sge} & cv,gg,u2,wd & 78 & 78 & 1 \\ 
J122740.86+513924.8 & GP CVn & 1571768231438756864 & 1.510810(2) & 18.74 & 390 & SU UMa & cv,gg,u1,u2,wd & 109 & 109 & 102 \\ 
J130514.73+582856.2 & SDSS J130514.73+582856.3 & 1578672472969644288 &  & 19.23 & 3911 & DN & cv,u2 & 115 & 115 &  \\
J133309.19+143706.9 & SDSS J133309.19+143706.9 & 3741755132951336192 & \textcolor{blue}{2.1148(1)} & 19.55 & 2028 & Polar & cv & 45 & 45 & 1 \\ 
J133941.08+484727.6 & V0355 UMa & 1558322303741820928 & 1.3754(4) & 17.37 & 150 & WZ Sge & cv,wd & 53 & 53 & 10 \\ 
J134052.05+151340.9 & CSS 100531:134052+151341 & 3742038776886730368 & 2.45 & 17.94 & 822 & U Gem & cv,u2 & 23 &  &  \\
J135218.95+280917.1 & CSS 090705:135219+280917 & 1451958950484177280 &  & 20.1 & 2263 & U Gem & cv & 43 & 43 &  \\
J135642.39+613024.4 & V0427 UMa & 1661194359687172480 &  & 20.47 & 2330 & SU UMa: & cv & 18 & 1 &  \\
J142953.55+073231.3 & SDSS J142953.56+073231.2 & 1172464842754767488 &  & 20.23 & 1348 & U Gem & cv,u2 & 13 & 1 &  \\
J143317.77+101122.7 & SDSS J143317.78+101123.3 & 1176468611268115200 & 1.30177630(5) & 18.55 & 232 & WZ Sge & cv,gg,u1,u2,wd & 31 & 31 & 102 \\ 
\textcolor{blue}{J143435.36+334050.1} &  & 1287753107989718784 &  & 20.03 & 363 & \textcolor{blue}{WZ Sge:\S :} & u2 & 1 & 1 &  \\
J150240.90+333423.4 & NZ Boo & 1289860214647954816 & 1.413831(1) & 17.26 & 184 & SU UMa+E & cv,gg,u1,u2,wd & 109 & 109 & 84 \\ 
J160809.54+542131.4 & ELAISN1 MOS15-02 & 1429001113336695168 & \textcolor{blue}{1.4646(5)} & 19.43 & 1168 & \textcolor{blue}{ER UMa} & cv & 29 & 1 & 1 \\ 
J161935.78+524631.7 & DDE 32 & 1425200578380530048 & 1.675066(5) & 19.28 & 449 & Polar & cv & 25 & 25 & 25 \\ 
J163722.21-001957.1 & V2690 Oph & 4383291477376259968 & 1.61736(3) & 19.99 & 1254 & SU UMa & cv,u2 & 54 & 54 & 33 \\ 
J165127.25-131809.2 & SSS 130510:165127-131809 & 4332792802397772672 &  & 19.91 &  & U Gem & cv & 110 & 1 &  \\
J165244.83+333925.5 & SDSS J165244.84+333925.4 & 1314596271336013824 & \textcolor{blue}{1.67:} & 20.74 & 3281 & SU UMa & cv,u2 & 53 & 53 & 1 \\ 
J165930.37-125327.2 & V0841 Oph & 4333061392472253440 & 14.4313(2) & 13.65 & 815 & Nova & cv & 70 & 70 & 80 \\ 
J170515.34+724403.3 & MASTER OT J170515.37+724402.5 & 1655084339211334016 &  & 19.41 & 2686 & \textcolor{blue}{SU UMa} & cv & 101 & 1 &  \\
J171752.03-070654.2 & MASTER OT J171752.02-070654.6 & 4360435241974939392 &  & 20.52 & 3089 & U Gem & cv & 19 & 1 &  \\
J172148.84-051713.2 & CSS 120611:172149-051713 & 4362925292215906176 &  & 20.12 & 3942 & U Gem & cv & 36 & 1 &  \\
\textcolor{blue}{J172805.34+775117.4} &  & 1705302990918074880 &  & 20.62 & 974 & \textcolor{blue}{WZ Sge:} & u2 & 1 & 1 &  \\
J173924.38+050049.7 & ASASSN-15mk & 4473606466292898688 &  & 20.54 &  & DN & cv & 24 & 1 &  \\
J174045.86+060350.5 & PBC J1740.7+0603 & 4485786821746343680 & 1.6826(50) & 15.84 & 232 & \textcolor{blue}{Polar: } & cv & 50 & 50 & 50 \\ 
J174056.31+025831.6 & MG1 995493 & 4376325246582929920 & \textcolor{blue}{4.143(3):} & 17.81 & 928 & \textcolor{blue}{U Gem\ddag } & cv,gg,u1,u2 & 34 & 1 & 1 \\ 
J174759.88+032859.7 & ASASSN-15fm & 4472400370757130496 &  & 20.07 & 957 & DN & cv & 30 & 1 &  \\
\textcolor{blue}{J181111.75+124301.7} &  & 4496951052360182656 &  & 19.79 & 689 & \textcolor{blue}{SU UMa:} & wd & 1 & 1 &  \\
J182122.51+614855.2 & ASASSN-13at & 2159834324676298880 & 1.901(5) & 19.46 & 764 & SU UMa & cv,gg,u1,u2 & 118 & 32 & 32 \\ 
J182138.61+615904.0 & ASASSN-14gg & 2159842815827966720 & 1.44(3)* & 19.78 & 722 & SU UMa & cv,wd & 68 & 1 & 117 \\ 
J182205.84+270849.8 & MASTER OT J182205.85+270849.7 & 4585224931075337216 &  & 20.66 & 1772 & \textcolor{blue}{SU UMa} & cv & 91 & 1 &  \\
J182643.55+613816.5 &  & 2159773512236202624 & \textcolor{blue}{2.0654} & 20.62 & 2118 & \textcolor{blue}{SU UMa} & cv & 11 & 1 & 1 \\ 
J183052.72+265514.9 & MGAB-V733 & 4585376839774011520 &  & 18.79 & 1175 & \textcolor{blue}{SU UMa} & u2 & 9 & 1 &  \\
J183211.36+615505.9 & ASASSN-13ah & 2159698023890909184 & 1.5874(3)* & 20.47 & 960 & SU UMa & cv & 89 &  & 99 \\ 
J192914.75+202759.5 & NQ Vul & 2017742684676480896 & 3.516(2) & 17.29 & 1170 & Nova & cv & 83 & 83 & 65 \\ 
J192927.83+202035.0 & 1RXS J192926.6+202038 & 2017736641707652608 & 3.89(3) & 17.58 & 567 & U Gem & cv & 120 & 112 & 112 \\ 
J195837.08+164128.4 & AW Sge & 1820137772050526464 & 1.788(5)* & 19.29 & 604 & SU UMa & cv & 58 & 1 & 123 \\ 
J200318.50+170251.4 & RZ Sge & 1820209309009288960 & 1.63872 & 18.29 &  & SU UMa & cv & 58 & 58 & 14 \\ 
J200611.52+334237.6 & V1363 Cyg & 2058291887543939968 & 2.42(14): & 16.28 & 1641 & U Gem & cv & 82 & 116 & 27 \\ 
J200934.41+550525.7 &  & 2185841249306525696 &  & 16.52 & 1707 & \textcolor{blue}{NL} & gg & 49 & 1 &  \\
J202812.47+414836.3 & V2467 Cyg & 2068080976989664768 & 3.83 & 18.04 & 1847 & IP: & cv & 20 & 76 &  \\
J204638.68+603803.0 & V0713 Cep & 2194206707428148480 & 2.0500424(1) & 18.35 & 319 & SU UMa+E & cv & 52 & 1 & 57 \\ 
J205423.76+601706.7 & V0962 Cep & 2194035978188297728 &  & 19.28 & 3095 & \textcolor{blue}{Nova} & cv & 103 & 1 &  \\
J212654.54-012054.3 & CRTS J212654.5-012054 & 2686013551247649152 &  & 19.06 & 1676 & \textcolor{blue}{Polar} & cv,u2 & 36 & 36 &  \\
J213432.34-012040.5 & CSS 081120:213432-012040 & 2686614060690086016 &  & 18.59 &  & \textcolor{blue}{U Gem} & cv & 36 & 1 &  \\
J221832.75+192520.2 & Swift J2218.5+1925 & 1777232423131443840 & 2.159(24) & 17.31 & 240 & Polar & cv,gg,u1,u2 & 22 & 62 & 62 \\ 
J221900.24+201831.2 & SDSS J221900.24+201831.1 & 1778468682222798080 & 3.36(1) & 18.23 & 821 & DN & gg,u1,u2 & 8 & 8 & 8 \\ 
J222117.33+194817.6 & ASASSN-19wi & 1777640754262037248 &  & 18.77 & 708 & \textcolor{blue}{DN} & gg,u1,u2 & 24 & 1 &  \\
J234440.54-001206.1 & MLS 100904:234441-001206 & 2642699218384434432 & 1.8411(7)* & 20.07 & 668 & SU UMa & cv,u2 & 110 & 43 & 66 \\

 \end{longtable}
\noindent
Cartons:\, cv\,\textcolor{blue}{mwm\_cb\_cvcandidates}, gg\,\textcolor{blue}{mwm\_cb\_gaiagalex}, u1\,\textcolor{blue}{mwm\_cb\_uvex1}, u2\,\textcolor{blue}{mwm\_cb\_uvex2}, u4\,\textcolor{blue}{mwm\_cb\_uvex4}, wd\,\textcolor{blue}{mwm\_wd}

\noindent
References:\, 1\,\textcolor{blue}{This work}, 2\,\citet{1997PASP..109.1359T}, 3\,\citet{2012ATel.4516....1W}, 4\,\citet{1988AN....309...91R}, 5\,\citet{2013PASP..125..506T}, 6\,\citet{2023ApJ...945..141R}, 7\,\citet{1990IAUC.5024....2M}, 8\,\citet{2023MNRAS.524.4867I}, 9\,\textcolor{blue}{MGAB Variable Star Catalog}, 10\,\citet{2006MNRAS.365..969G}, 11\,\citet{2016ATel.8956....1M}, 12\,\citet{2018AJ....155...28S}, 13\,\citet{2010MNRAS.402..436W}, 14\,\citet{2014PASJ...66...90K}, 15\,\textcolor{blue}{Lamost 8/1/2016}, 16\,\citet{1985AJ.....90.1837S}, 17\,\citet{1960JO.....43...17P}, 18\,\citet{2013ATel.5118....1S}, 19\,\citet{2017ATel10415....1V}, 20\,\citet{2007IAUC.8821....1N}, 21\,\citet{1970MitVS...5...99R}, 22\,\citet{2009ATel.2177....1T}, 23\,\citet{2012AJ....144...81T}, 24\,\textcolor{blue}{ASAS-SN CV candidate}, 25\,\citet{2016arXiv160908511D}, 26\,\textcolor{blue}{vsnet 19297}, 27\,\textcolor{blue}{vsnet 13294}, 28\,\textcolor{blue}{VSNET 12318}, 29\,\citet{2013ApJ...766...60G}, 30\,\citet{2019MNRAS.486.2422P}, 31\,\citet{2007AJ....134..185S}, 32\,\citet{2020AJ....160....6T}, 33\,\citet{2008MNRAS.391..591S}, 34\,\citet{2007AJ....134.1488K}, 35\,\citet{2015AJ....149..128T}, 36\,\citet{2014MNRAS.441.1186D}, 37\,\citet{2012ATel.4441....1D}, 38\,\citet{2020TNSTR.768....1H}, 39\,\citet{2021A&A...648A..44M}, 40\,\citet{2017TNSTR.925....1D}, 41\,\citet{2014PASJ...66...30K}, 42\,\citet{2019MNRAS.487.1120S}, 43\,\citet{2014MNRAS.443.3174B},  45\,\citet{2009AJ....137.4011S}, 46\,\textcolor{blue}{VSNET 19440}, 47\,\citet{2016ATel.9862....1S}, 48\,\citet{2005ApJ...620..929H}, 49\,\citet{2018AJ....156..241H}, 50\,\citet{2015AJ....150..170H}, 51\,\citet{1992A&A...259..198J}, 52\,\citet{2003IBVS.5461....1A}, 53\,\citet{2005AJ....129.2386S}, 54\,\citet{2002AJ....123..430S}, 55\,\citet{1987A&AS...70..481B}, 56\,\citet{2005AJ....130..825W}, 57\,\citet{2011JBAA..121..233B}, 58\,\citet{1982A&AS...48..383V}, 59\,\citet{2014CBET.4013....1M}, 60\,\citet{2019TNSTR.241....1D}, 61\,\citet{2020JAVSO..48...53S}, 62\,\citet{2013AJ....146..107T}, 63\,\citet{2013ATel.5536....1Y}, 64\,\citet{2003A&A...404..301R}, 65\,\citet{2021RNAAS...5..150S}, 66\,\citet{2010PASJ...62.1525K}, 67\,\textcolor{blue}{vsnet 15408}, 68\,\citet{2018TNSTR1819....1D}, 69\,\citet{1997PASP..109..477T}, 70\,\citet{1983ApJS...53..523W}, 71\,\citet{2017PASJ...69...75K}, 72\,\citet{2020ApJS..249...18C}, 73\,\citet{1997PASP..109.1241T}, 74\,\citet{2010A&A...524A..86S}, 75\,\citet{2020AJ....159...43H}, 76\,\citet{2009AN....330...77P}, 77\,\citet{1966AN....289....1H}, 78\,\citet{2011AJ....142..181S}, 79\,\citet{2018MNRAS.477.3145J}, 80\,\citet{2006PASP..118..687P}, 81\,\citet{2010ASPC..435..297S}, 82\,\citet{1971RA......8..167M}, 83\,\citet{1977A&A....55..307C}, 84\,\textcolor{blue}{Inight et al in prep}, 85\,\citet{2013MNRAS.429.2143C}, 86\,\citet{2020AJ....159..198S}, 87\,\citet{2007PASP..119..494S}, 88\,\citet{1997AcApS..17..107W}, 89\,\citet{2013ATel.5052....1S}, 90\,\citet{2000ApJS..128..387L}, 91\,\citet{2014ATel.6049....1D}, 92\,\citet{1994ApJ...420..830S}, 93\,\citet{2018TNSTR.423....1D}, 94\,\citet{2006MNRAS.373..687S}, 95\,\citet{2018A&A...620A.141R}, 96\,\citet{2009PASJ...61S.395K}, 97\,\citet{1989A&AS...78..145B}, 98\,\textcolor{blue}{VSNET 7967}, 99\,\citet{2016PASJ...68...65K}, 100\,\citet{2017TNSTR.547....1D}, 101\,\citet{2013ATel.5676....1B}, 102\,\citet{2008MNRAS.388.1582L}, 103\,\citet{2015MNRAS.454.1297S}, 104\,\citet{1995ApJ...440..336C}, 105\,\citet{2017ApJS..228...19C}, 106\,\citet{2009ApJ...696..870D}, 107\,\citet{2018ATel11512....1F}, 108\,\citet{2009ASPC..404..272A}, 109\,\citet{2006AJ....131..973S}, 110\,\textcolor{blue}{CRTS candidate CV}, 111\,\citet{2017TNSTR.933....1D}, 112\,\citet{2017RNAAS...1...29T}, 113\,\citet{2003AJ....126.1499S}, 114\,\citet{2020AJ....159..114O}, 115\,\citet{2004AJ....128.1882S}, 116\,\citet{1992A&AS...93..419B}, 117\,\textcolor{blue}{vsnet 20079}, 118\,\citet{2013ATel.5178....1G}, 119\,\citet{2005ApJ...620..422S}, 120\,\citet{2011AstL...37...91D}, 121\,\citet{2015PASJ...67..105K}, 122\,\citet{1996PASP..108..894T}, 123\,\citet{2008JBAA..118..145S}, 124\,\citet{2009ATel.2086....1D} 
\end{landscape}
\twocolumn
\clearpage

\section{Spectral confirmation of previously known systems}\label{section:prevknown}

The spectra and light curves of the following objects are shown in the supplementary material. We note below any new information; we have not commented on well-studied systems where we have nothing new to add.


\subsection{J0213+5338}
The spectrum shows Balmer emission lines. Together with the frequent outbursts and superoutbursts in the ZTF light curve this is consistent with the previous classification of SU\,UMa and a period of 2.549\,h. 

\subsection{J0221+7322}
The white dwarf is visible in the spectrum and there is a double peaked H$\alpha$ line. This combination is confirmation of the WZ\,Sge classification. 

\subsection{J0241+5936}
The spectrum is reddened by $0.39\,\mathrm{mag}$ due to being in the Galactic plane. It shows Balmer emission lines. There is an outburst in the ZTF light curve and this, together with its position in the HR diagram, confirms that this is a U\,Gem. 

\subsection{J0431+3527}
Although the spectrum of this object has a relatively low SNR, it reveals a strong double-peaked H$\alpha$ emission line. There are four outbursts apparent in the ZTF light curve including at least one superoutburst~--~confirming the existing classification of SU\,UMa.

\subsection{J0505+2225}
The spectrum shows Balmer and \ion{He}{i} emission lines. The ZTF light curve shows dwarf nova outbursts confirming the previous classification of SU\,UMa.

\subsection{J0524--0705}
Despite poor flux calibration, the Balmer, \ion{He}{ii} and \ion{He}{i} emission lines are visible in the spectrum. The ZTF light curve shows seven outbursts two of which appear to be superoutbursts with superhumps visible. We classify this as a probable SU\,UMa.

\subsection{J0612+2528}
This is HQ\,Gem and is  positioned on the edge of the main sequence in the HR diagram. The spectrum shows Balmer and \ion{He}{i} emission lines.  The light curve shows multiple outbursts, confirming that this is a U\,Gem.

\subsection{J0621+4042}
This is  positioned on the edge of the main sequence in the HR diagram. Lines from the donor are visible in the spectrum as well as Balmer emission lines which confirm the classification of U\,Gem.

\subsection{J0624+0021}\label{sec:j0624}
\begin{figure} 
\includegraphics[width=\columnwidth]{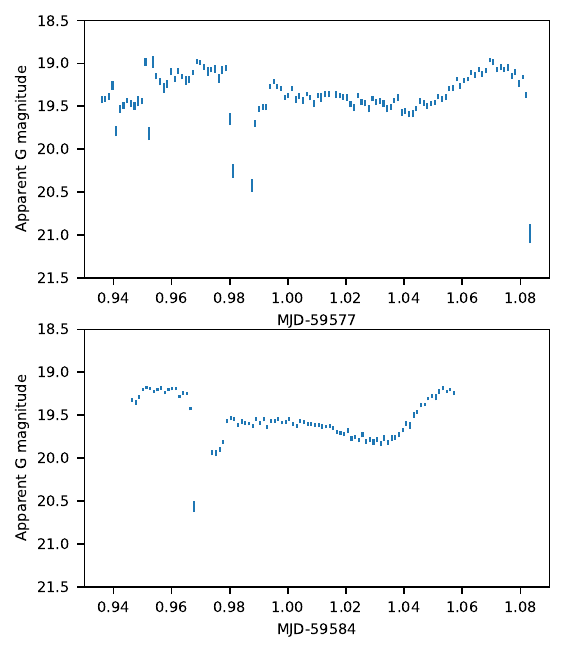}
\caption{\label{fig:LT_telescope} Photometric observations of J0624+0021 using the Liverpool Telescope. Top panel: 2021 December 02 showing a clear eclipse and the start of a second. Bottom panel: 2022 January 05 showing a single eclipse.   }   
\end{figure}

We obtained photometric light curves (Fig.\,\ref{fig:LT_telescope}) which show a deep eclipse. These curves were combined with ZTF data to determine the correct alias resulting in the ephemeris $\mathrm{HJD}(\phi=0)\,  2\,459\,578.48972(7) + N \times 0.101240(1)$.  Lines from the donor are visible in the spectrum, along with Balmer emission lines from the accretion disc. The ZTF light curves show outbursts and at least one superoutburst causing us to classify this as a SU\,UMa in the period gap.

\subsection{J0626--0355}
The spectrum shows Balmer, \ion{He}{ii} and \ion{He}{i} emission lines which, together with the outbursts in the ZTF light curves, confirm the classification as a dwarf nova.

\subsection{J0629--0335}
The spectrum shows Balmer emission lines albeit with poor calibration which confirm the classification in  \citet{2009ASPC..404..272A} of a Z\,Cam. 

\subsection{J0630+0239} 
The spectrum shows single-peaked Balmer emission lines as well as absorption lines from the donor although the flux calibration is poor. These characteristics are  consistent with the classification of \citet{2017PASJ...69...75K} that this is a low-inclination SU\,UMa.

\subsection{J0633+0324}

Double-peaked Balmer emission lines from an accretion disc as well as absorption lines from the donor are visible in the spectrum. It is very close to the main sequence in the HR diagram and the SED indicates a mid-K type donor. The donor characteristics together with the outbursts visible in the ZTF light curves lead us to the classification of U~Gem.

\subsection{J0708--1245}
This is ASASSN-16an. It is close to the galactic plane ($b=-2.1$) and significantly reddened with $E(B-V)\simeq0.7$. The spectrum shows  Balmer and \ion{He}{i} emission lines together with absorption lines from the donor. The ZTF light curve shows multiple outbursts. Taken together this confirms the U~Gem classification.
 
\subsection{J0737+2055} 
 
The spectrum shows narrow  Balmer and \ion{He}{i} emission lines indicating a low inclination system. There is a single five-magnitude outburst in the ZTF light curve, however there are too few data points to confirm whether or not this is a superoutburst. It is closer to the white dwarf cooling sequence than the main sequence in the HR diagram. This is probably a SU\,UMa.

\subsection{0751+3006}
The spectrum shows strong  Balmer and \ion{He}{i} emission lines. The spectrum together with the light curve outbursts and known period of 1.3918\,h confirm the SU\,UMa classification.

\subsection{J0752+2943}
The spectrum shows Balmer and \ion{He}{i} emission lines and also emission of \ion{He}{ii} at 4686\,\AA. The CRTS light curve shows a  $\sim3$ magnitude variation over a five year period with a 1.5 magnitude outburst on  $\mathrm{MJD}=55\,211$ whilst the ZTF light curve shows a 1.5\,mag outburst on $\mathrm{MJD}=58\,511$. Curiously the long term light curve shows a smooth change not typical of a high-low switch.
We checked nearby objects in CRTS and they did not show this type of variation. 
There is ellipsoidal modulation in the ZTF light curve at a period of 1.1445(7)\,h implying an orbital period of 2.289(2)\,h.
Together with the outbursts this confirms the classification of a dwarf nova.

\subsection{J0753+2943}
The spectrum shows strong Balmer and  \ion{He}{i} emission lines and J0753+2943 is located between the main sequence and the white dwarf cooling sequence in the HR diagram. There are two outbursts visible in the ZTF light curve. A periodogram of the ZTF light curve shows a reliable period of 1.4455(1)\,h.

\subsection{J0755+2646}
\begin{figure} 
\includegraphics[width=0.99\columnwidth]{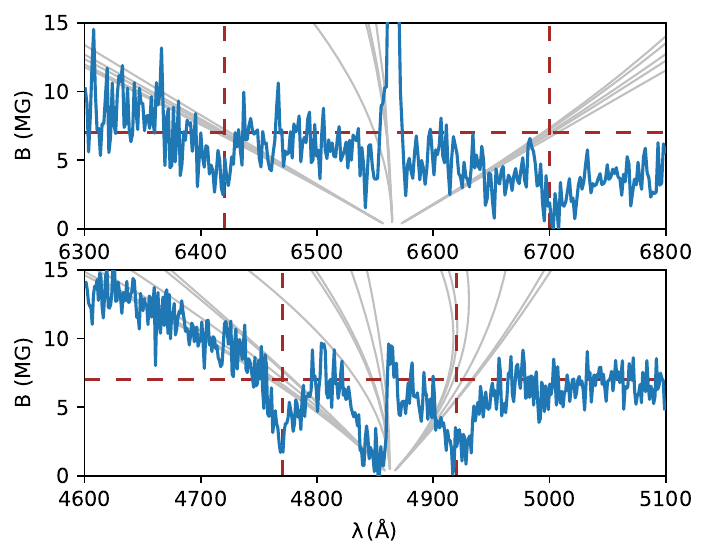}
\caption{\label{fig:755 zeeman} Spectra of H$\alpha$ and H$\beta$ lines of J0755+2646 showing the Zeeman splitting lines \citep{2014ApJS..212...26S}. The lines coincide with absorption lines for a magnetic field of $B\simeq7$\,MG. }
\end{figure}

We have 25 radial velocity measurements with moderate SNR over five epochs spanning 86 days. We find a period of 1.4883(1)\,h. The spectrum shows Zeeman splitting around weak H$\alpha$, H$\beta$ and H$\gamma$ lines (see Fig.\,\ref{fig:755 zeeman}).  This is therefore a magnetic white dwarf and not a U\,Gem dwarf nova; the field strength of $B\simeq7$\,MG suggests an intermediate polar although polars can have magnetic fields as weak as this \citep{1995MNRAS.273...17F}.  There is slight contamination in the spectrum from a background galaxy. It is currently in a low state (\textit{Gaia} EDR3 $m_\mathrm{G}=18.74$) but previously it has been in a high state (Pan-STARRS $m_{\mathrm{g}}=16.6$ and \textit{Gaia} DR2 $m_\mathrm{G}=16.82$).  The transient reported by \citet{2014CBET.4013....1M} is the same magnitude as the high states reported by Pan-STARRS and \textit{Gaia} DR2. The magnitude reported by \textit{Gaia} DR2 is based on 160 observations which would suggest a prolonged high state rather than transient behaviour.  This system might also be a low accretion rate polar with the white dwarf accreting material from the donor's wind rather than Roche-lobe overflow \citep{2012MNRAS.423.1437B}.
We tentatively classify this as a polar.

\subsection{J0805+3540}

The spectrum shows Balmer, \ion{He} {ii}  and \ion{He}{i} emission lines. The spectrum has the appearance of a K-star while an absorption dip just over 5000\,\AA\  is reminiscent of QZ\,Per \citep{2002PASP..114.1117T}. A periodogram of the ZTF light curve shows ellipsoidal modulation with an orbital period of 5.2\,h. This is a U\,Gem and may have an evolved donor.

\subsection{J0902--0142}
The spectrum shows double peaked Balmer and \ion{He}{i} emission lines. There are hints of TiO absorption lines from the donor $-$ consistent with the spectral energy distribution.  We have a tentative period from the radial velocities of 2.04(10)\,h. The ZTF light curve shows outbursts; taken together with the tentative period this is probably a SU\,UMa.

\subsection{J0903--0133}
The spectrum shows strong Helium emission and no hydrogen lines. This confirms that this object is an AM\,CVn as suggested by \citet{2014MNRAS.439.2848C}.

\subsection{J0926+0345}
J0926+0345 was in outburst during one of the spectra (see Fig.\,13 in the supplementary material).  The quiescent spectrum shows strong Balmer and \ion{He}{i} emission lines. The ZTF light curve shows numerous outbursts and at least one superoutburst. It is nearer the white dwarf cooling sequence than the main sequence in the HR diagram. We therefore classify this as a SU\,UMa.

\subsection{J0932+0343}

The spectrum shows strong Balmer and \ion{He}{i} emission lines. TiO and Na absorption lines from the donor are visible and also absorption lines from the white dwarf  indicating a low accretion rate. The ZTF light curve shows frequent outbursts with the hint of a superoutburst. This is probably a SU\,UMa.

\subsection{J1356+6130}

The spectrum shows weak double peaked Balmer and \ion {He}{i} emission lines. The white dwarf and the donor are both visible in the spectrum indicating a low accretion rate. The photometry is dominated in the red by the bright donor. The ZTF light curve shows six outbursts including a superoutburst. This confirms that this is a SU\,UMa.  

\subsection{J1429+0732}

The spectrum shows double peaked Balmer and \ion {He}{i} emission lines. The  donor is very visible. The ZTF light curve shows ten outbursts consistent with the classification of U\,Gem. 

\subsection{J1608+5421}

We have twelve usable  radial velocities in three epochs. The tallest peak in the \textsc{The Joker} periodogram is at $\simeq1.4646$\,h and the amplitude of the  periodogram falls off sharply after the two adjacent aliases. The ZTF light curve shows at least ten outbursts (actually probably superoutbursts) with  a quiescent magnitude $m\simeq22.5$. Analysis of the light curve between 19 and 20 magnitude and \mbox{$58\,376<\mathrm{MJD}<58\,390$} shows superhumps with a period of $\simeq1.506$\,h which confirms that $\simeq1.4646$\,h is the correct alias.   We therefore estimate the period to be 1.4646(5)\,h. The frequent superhumps in the light curve suggest that this is a ER\,UMa type system \citep{1995PASP..107..443R,2013arXiv1301.3202K}~--~a sub-type of SU\,UMa.

\subsection{J1651--1318}
Balmer emission lines are visible despite poor SNR and flux calibration.
The ZTF light curve shows four outbursts whilst the quiescent magnitude is at least 21 and so the \textit{Gaia} and Pan-STARRS magnitudes (19.9 and 19.5 respectively) are based on the magnitude of the outbursts. The outbursts and emission lines confirm the classification of U~Gem.

\subsection{J1705+7244}
The spectrum shows strong Balmer  and \ion{He}{i} emission lines with neither the donor nor the white dwarf visible. The ZTF light curve shows frequent outbursts and at least two superoutbursts. We therefore classify this as a SU\,UMa. 

\subsection{J1717--0706}
The spectrum shows Balmer  and \ion{He}{i} emission lines with neither the donor nor the white dwarf visible. It is next to the main sequence on the HR diagram and the donor is clearly visible in the SED.  It is very faint ~--~\textit{Gaia} DR3 reports $m_\mathrm{G}=20.52$. There is no ZTF data although the CRTS data shows two outbursts reaching $m\simeq16.5$. 
The emission lines and outbursts confirm that this is a U~Gem.

\subsection{J1721--0517}
The spectrum shows strong Balmer, Paschen  and \ion{He}{i} emission lines with a very red continuum from the donor. It is close to the main sequence in the HR diagram. The ZTF light curve shows three outbursts of $\Delta m \simeq4.5$. We therefore confirm that it is a U\,Gem. 

\subsection{J1739+0500}
The spectrum shows Balmer  and \ion{He}{i} emission lines with neither the donor nor the white dwarf visible. The system is faint ($m_\mathrm{G}=20.48$) and the spectrum has a poor SNR. The ZTF light curves show three outbursts and we conclude that this is a dwarf nova.

\subsection{J1740+0258}\label{sec:j1740}
\begin{figure} 
\includegraphics[width=\columnwidth]{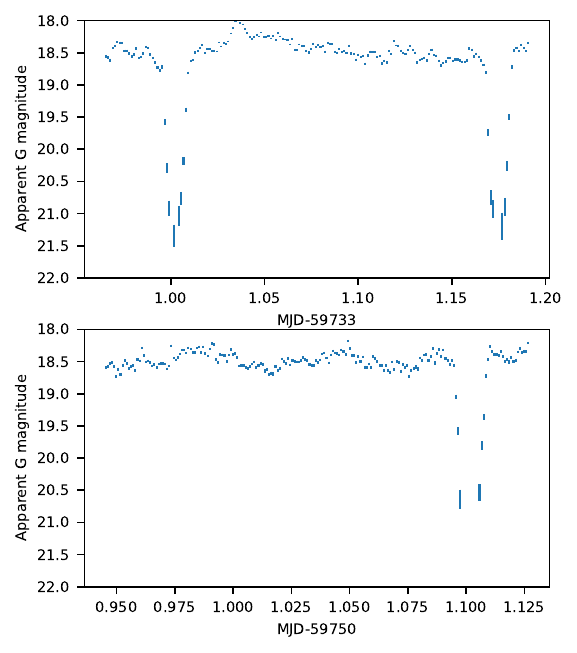}
\caption{\label{fig:LT_telescope_2} Photometric observations of J1740+0258 using the Liverpool Telescope. Top panel: 2022 June 03 showing two clear eclipses. Bottom panel: 2022 June 20 showing a single eclipse.   }   
\end{figure}

The spectrum shows Balmer and \ion{He}{i} emission lines. TiO and Na 8194\,\AA \ absorption lines from the donor are visible.  
The ZTF light curve showed two outbursts with unusual state changes and J1740+0258 also appeared to be eclipsing.  We therefore obtained LT photometry (see Fig.\,\ref{fig:LT_telescope_2}) which confirmed that J1740+0258 was eclipsing with ephemeris  $\mathrm{HJD}(\phi=0) 2\,459\,734.5074(2) + N \times 0.172724(2)$. This is therefore a U~Gem.

\subsection{J1747+0328}
The spectrum shows a CV in outburst with a synthetic magnitude $m_\mathrm{G}=17.5$ consistent with the outbursts in the ZTF light curve. This confirms the classification of dwarf nova.

\subsection{J1821+6159}
The spectrum shows strong Balmer, Paschen  and \ion{He}{i} emission lines with neither the donor nor the white dwarf visible. The ZTF light curve shows seven outbursts consistent with the published classification of SU\,UMa and superhump period of 1.44\,h.  

\subsection{J1822+2708}
The spectrum shows strong double-peaked Balmer, Paschen, \ion{He}{i} and Ca emission lines. It is midway between the main sequence and the white dwarf cooling sequence in the HR diagram. The ZTF light curve shows three outbursts including a superoutburst. This is a SU\,UMa.

\subsection{J1826+6138}

The spectrum shows weak double-peaked Balmer emission lines.  The ZTF light curve shows frequent outbursts and also five superoutbursts leading to a classification of SU\,UMa. We find a  period of 2.0654\,h from the ZTF periodogram. 

\subsection{J1830+2655}
We have both outburst and quiescent spectra (see Fig.\,13 in the supplementary material). 
The quiescent spectrum shows strong Balmer, Paschen  and \ion{He}{i} emission lines with neither the donor nor the white dwarf visible. The ZTF light curve shows frequent short outbursts and at least one superoutburst resulting in our classification of SU~UMa. The light curve is unusual as the quiescent magnitude appears to fall by over a magnitude between $\mathrm{MJD}=58800$ and $\mathrm{MJD}=59000$. 

\subsection{J1958+1641}
The spectrum shows double-peaked Balmer emission lines. The ZTF light curve shows seven outbursts including at least one superoutburst. A periodogram of the ZTF light curve yields an orbital period of 1.723023(5)\,h  which is consistent with the published superhump period of 1.735(16)\,h.

\subsection{J2009+5505}

The spectrum appears to be either that of a novalike or a dwarf nova in outburst. However the light curve shows no signs of an outburst and its position in the \textit{Gaia} HR diagram is consistent with a novalike. This object was classified as a subdwarf in \citet{2017A&A...600A..50G} however the spectrum shows H$\alpha$ and H$\beta$ emission lines and that, together with the position on the HR diagram leads us to classify it as a novalike. 
The light curve also shows an unusual long term trend of increasing magnitude. 

\subsection{J2054+6017}
This is a classical nova that erupted in 2014 \citep{2015MNRAS.454.1297S}. It is now very faint (synthetic magnitude $m_\mathrm{G}=20.6$) leading to the poor SNR in the spectrum, inconsistent with the \textit{Gaia} magnitude ($m_\mathrm{G}=19.3$).

\subsection{J2134--0120}
This system was identified as a transient by CRTS (CSS 081120:213432$-$012040) and is classified as a U\,Gem by VSX. \citet{2014MNRAS.441.1186D} considered it a CV based on its CRTS light curve and photometry, however, \citet{2016MNRAS.456.4441C} decided against a dwarf nova classification based on the same CRTS light curve. The archival photometry available shows large-amplitude ($\simeq2$\,mag) variability on time scales of months to years, with no clear periodicity. When non-detections are accounted for the ZTF photometry appears to show the outbursts of a dwarf nova. The system is very blue, and also varies in the far-ultraviolet, with three separate \textit{GALEX} detections ($m_\mathrm{FUV}=18.6, 20.0, 23.5$). The \SDSSV\  spectrum reveals a blue continuum with weak Balmer absorption lines, and no noticeable emission lines. The spectrum was obtained close to peak brightness and is consistent with a dwarf nova in outburst. \textit{Gaia} has not obtained a parallax for this object and hence we cannot position it on the HR diagram. We therefore classify it as a U\,Gem. 

\subsection{J2221+1948}
The spectrum shows strong Balmer, Paschen, Ca (H and K)  and \ion{He}{i} emission lines with neither the donor nor the white dwarf visible. The ZTF light curve does not show outbursts and the CRTS light curve shows only one outburst from which we make a classification of dwarf nova.

\section{Well known systems}\label{section:wellknown}

The following well-studied systems have previously published spectra. We include their spectra and light curves in the supplementary material and note below  any new information.

\subsection{J0142+0017}
We obtained 16 high SNR  radial velocities over 3 epochs. The sine fit yielded an estimated period of 1.8777(1)\,h whilst \textsc{The Joker} identified 1.914\,h.  We therefore estimate the period as 1.88(4)\,h.  This is consistent with the previous classification of a polar. 

\subsection{J0155+0028}
This is FL\,Cet $-$ a known eclipsing polar.
This spectrum is that of a polar in a low state exhibiting cyclotron humps and Zeeman splitting.   It contrasts with a previous SDSS spectrum \citep{2002AJ....123..430S}  when it was in a high state. 

\subsection{J0207--0141}
We obtained 18 high SNR radial velocities over 3 epochs.  A sine fit yields an estimated period of 1.5515(1)\,h. \textsc{The Joker} identified 1.400\,h.  We therefore estimate the period as 1.55(15)\,h. The spectrum shows strong Balmer, \ion{He}{ii} and \ion{He}{i} emission lines consistent with the previous classification of a polar.

\subsection{J0233+0050}
The spectrum and ZTF light curve confirm the previous classification of WZ~Sge. However the long period (1.6\,h) and absence of the donor in the spectrum and SED indicate that this is probably a period bouncer.

\subsection{J0418+5107}\label{sec:NSPER}

Our MDM campaign on this object started with the modular spectrograph in 2018 November, and continued with the Ohio State Multi-Object Spectrometer (OSMOS;\citealt  {2011PASP..123..187M}) in 2018 December and 2019 January.  We obtained 40 usable velocities of H$\alpha$ emission; the spectrum also showed a contribution from a late-type secondary, and in 33 spectra we were able to measure usable absorption-line velocities by cross-correlating portions of the spectrum with a zero-velocity late-type template spectrum. The periods from the emission velocities and absorption velocities agreed within the uncertainties, with a weighted mean of 6.296(1)\,h. 
This is consistent with the U~Gem classification $-$ a long period with frequent outbursts. There is some reddening due to its proximity to the Galactic Plane.

\subsection{J0426+3541}
Our spectrum shows double peaked Balmer and \ion{He}{i} emission lines. The ZTF light curve shows outbursts and, intriguingly, variations in the quiescent level. It is a non-magnetic system with high inclination consistent with the previous SU\,UMa classification.

\subsection{J0443+4721}
This is an old nova which erupted in 2018 whose decline can be seen in the ZTF light curve. Comparing the spectrum with the SED shows that J0443+4721 is still in a  high state compared with its historical luminosity. It has an unusually long period of 3.4118\,d \citep{2020A&A...639L..10M}.

\subsection{J0627+1613}

J0627+1613 was initially identified from a transient by Gaia Alerts \citep{2018TNSTR.423....1D}. ZTF light curves show two transients  $\Delta m\sim2$ each with a duration of $\simeq15$\,d. Unusually J0627+1613 is positioned slightly above the main sequence in the HR diagram. The SED is essentially a blackbody with a surface temperature $\simeq3000$\,K. The spectrum is consistent with this and suggests that this is a CV with a $\sim$\,K0 donor which would explain the position in the HR diagram. We confirmed that the orbital period, estimated using the ZTF photometry by \citet{2020ApJS..249...18C}, to be 10.428\,h, is correct. This period is also consistent with a CV having a high-mass donor at an early stage in its evolution. 

\subsection{J0738+2855}

 The spectrum shows  Balmer and \ion{He}{i} emission lines together with absorption lines from the donor. The ZTF light curve suggests that it is eclipsing; there are also at least seven outbursts. \citet{2003AJ....126.1499S} found a period of 2.1\,h which is not consistent with the donor presence in the spectrum~--~a short period system would have a dim donor. The ZTF periodogram shows a period of 5.555(1)\,h. The \textit{TESS} periodogram shows a period of 5.55(2)\,h .  

\subsection{J0753+3758}
The spectrum shows strong  Balmer, \ion{He} {ii}  and \ion{He}{i} lines.
The ZTF light curve shows variability but no outbursts. 
We have 21 radial velocities with high SNR over four epochs spanning 84 days from which we find a period of 1.5363(1)\,h. This period is consistent with \textsc{The Joker}. This is probably a magnetic CV as previously reported.

\subsection{J0805+3540}

There are three spectra; two in a quiescent state and one during outburst. The outburst state (shown in Fig. 5 in the supplementary material) differs from the quiescent spectra by having a very strong \ion{He}{ii} emission line. There is the suspicion of TiO absorption lines from the M-dwarf in the quiescent spectra. It is located towards the white dwarf sequence in the HR diagram. The ZTF light curves show outbursts and we derive a period of 5.2069(2)\,h from a ZTF periodogram.. This is  a U\,Gem  and likely to have an evolved donor which would explain the helium emission and the short period for a relatively high mass donor.

\subsection{J0808+3550}

We have two spectra, one in quiescence and one in outburst (see Fig.\,13 in the supplementary material). The previously published spectrum \citep{2012ATel.4516....1W} was taken during an outburst. Our quiescent spectrum has  strong double-peaked Balmer and \ion {He} {i}  emission lines. There are frequent outbursts visible in the ZTF light curves all consistent with the dwarf nova classification. 

\subsection{J0819+2133}

The spectrum shows strong  Balmer and \ion{He}{i} emission lines and also TiO absorption lines from the donor. However the unusually variable ZTF light curve shows apparent changes in state over a period of $\sim1$\,y indicating a potentially magnetic system. The synthetic magnitude of our spectrum is 21.5 and so our observation was taken during a low state. This is probably an intermediate polar.

\subsection{J0822+2313}
This was identified by \citet{2009AJ....137.4011S} as a CV based on a strong, broad H$\alpha$ emission line in five exposures taken between 2004 November 05 and 2004 November 08. We have 29 exposures taken between 2020 Nov 11 and 2021 March 08 which show an M-dwarf profile but no sign of H$\alpha$ emission. There are no ultraviolet or X-ray observations shown in Vizier nor evidence of outbursts in ZTF (although CRTS shows some transients). In the absence of other evidence we can only consider this to be a Candidate CV. 

\subsection{J0836+2121}
This is CC\,Cnc which has a known period of 1.764(12)\,h and is in the Ritter \& Kolb catalog \citep{2003A&A...404..301R}. The ZTF light curve shows short outbursts but also apparent changes of state of magnitude $\Delta m\simeq1.5$ . 

\subsection{J0844+0239}
This is V495\,Hya. We measured a period using radial velocities of 4.9708(6)\,h which is consistent with, and improves upon, the period of 4.968(17)\,h reported by  \citet{2015AJ....149..128T}.

\subsection{J0845+0339}
This is V498\,Hya. The spectrum shows double peaked Balmer emission lines. There are clear absorption lines from the white dwarf but no evidence of the donor. Assuming the \citet{2021AJ....161..147B} distance of 1635\,pc is correct then V498\,Hya is located above the white dwarf cooling sequence on the HR diagram; however this distance is not consistent with the parallax nor the earlier \citet{2018AJ....156...58B} estimate of 302\,pc both of which would place V498\,HYa on the white dwarf cooling sequence. The period is 1.4256\,h from \citet{2009PASJ...61S.395K} based on a seven magnitude  superoutburst in January 2008. The ZTF light curves do not show any outbursts. From the size of the 2008 outburst and the absence of subsequent observed outbursts we conclude that it is probably a WZ\,Sge and also a period bouncer.

\subsection{J0851+0308}
This is a well studied system called CT\,Hya. In passing we note that the ZTF light curve shows an apparent brightening over a period of $\simeq1$\,yr of $\Delta m\simeq0.5$.

\subsection{J0926+0105}
In addition to asymmetric  Balmer and \ion{He}{i} emission lines the SDSS spectrum contains \ion{He}{ii} at 4686\,\AA. Both the donor and the white dwarf are visible in the form of broad Balmer absorption lines and an upturn at the red end of the SDSS spectrum, respectively. Currently \textit{Gaia} EDR3  has $m_\mathrm{G}=19.4$, however the SED plot shows a sequence of detections several magnitudes brighter suggesting that there has been a state change at some point. The ZTF light curve shows considerable variability, and from it, we obtain a period that is consistent with our spectroscopic period of 1.47234(1)\,h. This is more accurate than the period of 1.48\,h reported by \citet{2023ApJ...945..141R}.

\subsection{J0932+0109}
This is a bright, distant (3145\,pc) object that was identified as a novalike or potential polar in \citep{2003AJ....126.1499S}. The spectrum shows  strong Balmer, \ion{He}{ii} and \ion{He}{i} emission lines. The emission limes are asymmetric which, together with the \ion{He}{ii} are consistent with being a polar.  The ZTF data shows variations of $\Delta m \sim 1$ over a six minute interval which we presume is due to cyclotron beaming.  We therefore classify this as a polar.

\subsection{J1042+3718}
The spectrum shows distinct signs of cyclotron humps consistent with the previous classification of a polar.
We have five high SNR radial velocities in one epoch. We estimate the period to be 1.82(4)\,h which is consistent with \textsc{The Joker}.

\subsection{J1123+4317}
\citep{2014MNRAS.443.3174B} obtained a spectrum and classified this as a dwarf nova.
Our ZTF light curve shows a single superoutburst with clearly visible superhumps. We therefore classify this as a SU\,UMa.

\subsection{J1219+2049}

We have 10 moderate SNR radial velocities in three epochs and estimate the period to be 1.3018(1)\,h.  
The white dwarf absorption lines are clearly visible in the spectrum implying that this is a low accretion rate CV~--~consistent with a short period. It is located in the white dwarf cooling sequence in the HR diagram and taking account of the period and low accretion rate we classify this as a WZ\,Sge.

\subsection{J1333+1437}

We have 13 RV data points in 3 epochs. We have improved upon the period of $2.2 \pm0.1$ \,h from \citet{2008PASP..120..160S}. Our best fit is 2.1148(1)\,h  which is is consistent with the earlier observation and takes advantage of the longer baseline (59 days) of our observations. It is also consistent with a periodogram from the ZTF light curve. The spectrum shows strong Balmer, \ion{He}{ii} and \ion{He}{i} emission lines. The ZTF light curve, though sparse, shows evidence of a low state  of $m\sim21$ and a high state of $m \sim 20 $.  This, together with the \ion{He}{ii} emission lines, are confirmation of the previous classification of a polar.

\subsection{J1652+3339}
The spectrum shows strong double-peaked  Balmer  and \ion{He}{i} emission lines with both the donor and the white dwarf visible. The ZTF and CRTS light curves show frequent outbursts and at least two superoutbursts. Analysing the radial velocities using \textsc{The Joker} suggests that the orbital period may be $\simeq1.67$\,h. This is an SU\,UMa,

\subsection{J1740+0258} 
We obtained a  period of 4.143(3)\,h from LT photometry (Fig.\,\ref{fig:LT_telescope_2} ). The ZTF light curve also shows a significant state change reminiscent of the unusual CVs identified by \citet{2023MNRAS.524.4867I}.

\subsection{J1740+0603} 
The spectrum shows strong Balmer, Paschen,  \ion{He}{ii}  and \ion{He}{i} emission lines with neither the donor nor the white dwarf visible. The emission lines are asymmetric and, combined with the \ion{He}{ii} emission lines, cause us to classify this a probable polar.

\subsection{J2006+3342}
The ZTF light curve shows two ``outbursts'' with very long (months) decay times. \citet{2015MNRAS.451.2863S} report that this CV is surrounded by a faint nova shell which may account for this unusual behaviour.

\subsection{J2126--0120}
The quiescent spectrum shows strong Balmer, Paschen, \ion{He}{ii} and \ion{He}{i} emission lines with the donor but not the white dwarf visible. The emission lines appear to be asymmetric and there are cyclotron humps and these, together with the \ion{He}{ii} emission imply that this is magnetic and we therefore classify this as a polar.

\subsection{J2219+2018}
The spectrum shows strong Balmer, Paschen, \ion{He}{ii} \,and \ion{He}{i}\, emission lines with the donor but not the white dwarf visible. The ZTF light curve shows frequent outbursts together with an apparent standstill between $\mathrm{MJD}\simeq58\,970$ and $\mathrm{MJD}\simeq59\,250$.  We therefore classify this as a dwarf nova.

\section{Unusable observations}

\begin{table}
 \caption{The twelve CVs and CV candidates listed below were observed by SDSS-V, however, the quality of their spectra was too low to draw any conclusions.}
\centering
\label{tab:unusable_observations}
\begin{tabular}{ll}
\hline
SDSS name & Alternative name \\ \hline
SDSS\,J021229.78+570519.4   & UW\,Per \\
SDSS\,J062632.18+161622.9   & Gaia17cxa \\
SDSS\,J074121.62+313821.0   & ASASSN-15dx \\
SDSS\,J074813.35+290512.0   & V434\,Gem \\
SDSS\,J085228.73+020102.5   & CSS\,160419:085229+020103 \\
SDSS\,J142936.25+322629.2   & CSS\,140607:142936+322630 \\
SDSS\,J185234.98$-$001842.4   & V1724\,Aql \\
SDSS\,J200130.23+184255.6   &  \\
SDSS\,J200214.34+313634.7   & ASASSN-17eo	 \\
SDSS\,J200504.93+322122.5   & V550 Cyg \\
SDSS\,J201222.34+325927.5   &  \\
SDSS\,J201649.48+382109.3   & V1377 Cyg \\\hline
\end{tabular}
\end{table}



\bsp	
\label{lastpage}
\end{document}